\documentclass[lettersize,journal]{IEEEtran}
\usepackage{amsmath,amsfonts}
\usepackage{algorithmic}
\usepackage{algorithm}
\usepackage{array}
\usepackage{xcolor}
\usepackage[caption=false,font=normalsize,labelfont=sf,textfont=sf]{subfig}
\usepackage{textcomp}
\usepackage{stfloats}
\usepackage{url}
\usepackage{verbatim}
\usepackage{graphicx}
\usepackage{subcaption} 
\usepackage{cite}
\usepackage[english]{babel}
\usepackage{amsthm}
\usepackage{amsmath}
\usepackage{amssymb}

\newtheorem{theorem}{Theorem}
\usepackage{multirow}
\usepackage{booktabs} 
\usepackage{tcolorbox}
\tcbuselibrary{listings,breakable}

\begin{document}

\title{EByFTVeS: Efficient Byzantine Fault Tolerant-based Verifiable Secret-sharing in Distributed Privacy-preserving Machine Learning}

\author{Zhen Li,
Zijian Zhang*,~\IEEEmembership{Senior Member,~IEEE,}
Wenjin Yang,
Pengbo Wang,
Zhaoqi Wang,
Yan Wu,
Xuyang Liu*,
Jing Sun,
\thanks{Zhen Li, Zijian Zhang, Wenjin Yang, Pengbo Wang, Zhaoqi Wang, Yan Wu are with the School of Cyberspace Science and Technology, Beijing Institute of Technology, Beijing, China (email:zhen.li@bit.edu.cn;zhangzijian@bit.edu.cn;wenjinyang@bit.edu.cn;
wangpengbo@bit.edu.cn; zhaoqi\_wang@bit.edu.cn; wuyan.bit@gmail.com).}
\thanks{Xuyang Liu is with the School of Cyberspace Science and Technology, Beijing Institute of Technology, Beijing, China, and the School of Computer Science, The University of Auckland, New Zealand (email: liuxuyang@bit.edu.cn).}
\thanks{Jing Sun is with the School of Computer Science, The University of Auckland, New Zealand (email: jing.sun@auckland.ac.nz).}
 \thanks{Corresponding author: Zijian Zhang, Xuyang Liu.}}

\markboth{Journal of \LaTeX\ Class Files,~Vol.~14, No.~8, August~2021}%
{Shell \MakeLowercase{\textit{et al.}}: A Sample Article Using IEEEtran.cls for IEEE Journals}


\maketitle

\begin{abstract}
In the realm of Distributed Privacy-preserving Machine Learning (DPML), Verifiable Secret Sharing (VSS) serves as a fundamental primitive for guaranteeing data integrity against malicious actors. Despite its utility, VSS-based DPML frameworks frequently struggle with issues related to view consistency and prohibitive resource consumption. While recent integrations of Byzantine Fault Tolerance (BFT) protocols have attempted to mitigate these consistency issues, we identify a residual vulnerability inherent in current designs. We demonstrate that adversaries can exploit the lack of strict synchronization in share provision through a strategy we term Adaptive Share Delay Provision (ASDP). This strategy enables a novel adversarial technique, the ASDP-based Customized Model Poisoning Attack (ACuMPA), which allows attackers to tailor malicious gradients for specific targets. We provide a rigorous theoretical foundation explaining the efficacy of ACuMPA against existing safeguards. To neutralize this threat, we introduce EByFTVeS, a robust framework that couples BFT consensus with verifiable secret sharing. By enforcing a strict temporal lock on share commitments via the consensus mechanism, EByFTVeS effectively precludes the delayed injection of poisoned shares. Our theoretical analysis confirms the protocol's properties regarding validity, liveness, consistency, and privacy. Furthermore, comprehensive empirical evaluations indicate that EByFTVeS achieves superior computational efficiency compared to state-of-the-art VSS alternatives.

\end{abstract}


\section{Introduction}
Secure Multiparty Computation (SMC) and threshold cryptography rely heavily on Verifiable Secret Sharing (VSS)~\cite{vss-original} to establish trust in distributed environments. By extending classical secret sharing with cryptographic commitments, VSS empowers participants to verify the correctness of received shares without revealing the underlying secret. This capability is indispensable for robust distributed systems, ensuring that a dealer cannot distribute invalid or conflicting shares to honest participants. Consequently, VSS has emerged as a cornerstone technology for ensuring data integrity and preventing sabotage in collaborative computational tasks.

The application of VSS within Distributed Privacy-preserving Machine Learning (DPML) has garnered significant attention due to its potential to secure sensitive data updates, such as gradient exchanges. For instance, recent works have leveraged VSS to secure patient records in crowdsourcing healthcare systems~\cite{ss-healthcare} and to deter collusion among computing parties~\cite{eurocrypt-vss}. However, the practical deployment of these schemes is often hindered by two pervasive challenges: the equivocation of commitments and the exorbitant overhead associated with cryptographic operations.

First, the issue of consistency remains a critical vulnerability. In the absence of a unified broadcast channel, a malicious dealer acts as a single point of failure by potentially sending conflicting commitments to different participants. Similarly, compromised participants may attempt to replay obsolete shares during the reconstruction phase. Such inconsistencies can severely disrupt the learning process, facilitating model poisoning attacks that prevent the global model from converging. Second, the computational complexity of VSS primitives—particularly those involving homomorphic polynomial commitments—imposes a severe performance penalty. The necessity of broadcasting validity proofs alongside raw shares exacerbates communication latency, rendering standard VSS protocols ill-suited for high-throughput DPML scenarios involving massive parameter sets~\cite{mp-spdz}. While some benchmarking frameworks have attempted to alleviate online costs by offloading Beaver triple generation to an offline phase, the online verification of secret consistency remains a bottleneck.

To address the consistency dilemma, researchers have begun incorporating Byzantine Fault Tolerance (BFT) mechanisms into VSS workflows~\cite{basuEfficientVerifiableSecret2019}. By mandating that dealers publish shares and commitments via a BFT-ordered ledger, these protocols aim to create an immutable record of the sharing phase. Theoretically, this prevents a dealer from equivocating or updating shares mid-round. However, we reveal that these measures are insufficient against a sophisticated adversary. We uncover a timing-based vulnerability: the Adaptive Share Delay Provision (ASDP). By strategically delaying the broadcast of shares until honest participants have revealed their contributions, a malicious dealer can craft and inject "customized" shares that bypass validity checks while poisoning the model.

In this work, we present a comprehensive solution to these challenges. Our contributions are fourfold:
\begin{itemize}
\item We formally define the Adaptive Share Delay Provision (ASDP) strategy. We demonstrate how a malicious dealer can leverage information asymmetry to generate arbitrary shares for victim participants, and we prove that these fabricated shares will inevitably be accepted by the victims under current protocol definitions.
\item We formulate the ASDP-based Customized Model Poisoning Attack (ACuMPA). By utilizing the ASDP strategy, we show how an adversary can circumvent standard defenses, such as cosine similarity filtering, to corrupt the global model aggregation process.
\item We propose EByFTVeS, an optimized VSS scheme integrated with BFT consensus. The core innovation of EByFTVeS lies in its consensus-driven synchronization, which mandates that only shares committed within a verified consensus window are eligible for aggregation. This mechanism effectively eliminates the time window required for the ACuMPA attack.
\item We provide a rigorous theoretical analysis of the proposed scheme, covering validity, liveness, consistency, and privacy. Empirical comparisons with existing state-of-the-art solutions confirm that EByFTVeS delivers significant improvements in both security and execution efficiency.
\end{itemize}

\section{Related Works}
Verifiable Secret Sharing (VSS) plays a fundamental role in constructing robust Secure Multiparty Computation (SMC) protocols, particularly when dealing with active adversaries who may deviate arbitrarily from the protocol \cite{vss-original,vss-feldman,vss-benaloh}. 
By ensuring that the initial secret-sharing phase is verifiable, VSS prevents a malicious party from distributing inconsistent or invalid shares that could compromise the entire computation. 

Related works in this area focus on designing efficient VSS schemes that can be seamlessly integrated into various SMC frameworks, minimizing the overhead while maximizing security guarantees against a variety of adversarial models. For instance, BEAR \cite{ssfl-jsac} pioneers the use of VSS to securely compute gradient distances for KRUM aggregation, leveraging polynomial operations to ensure robustness. Similarly, ELAS \cite{elas-beaver} incorporates secret sharing combined with Beaver triples to calculate the $l_2$ norm for achieving secure aggregation. Furthermore, RFLPA \cite{RPLPA-nips} introduces a novel dot-product-based aggregation rule, leveraging secret sharing to compute cosine similarity and effectively defend against poisoning attacks, showcasing the versatility of VSS in distributed machine learning scenarios.
Frankly, almost all of the aforementioned works have to apply Beaver triples to support multiplication operations. Since the construction of Beaver triples relies on multiple rounds of oblivious transfer, which is heavy to compute and communicate intermediate parameters, therefore, the efficiency of these works has yet to be a great challenge.

The synergy between VSS and Threshold Cryptography is crucial for enabling distributed cryptographic operations where a collective of parties can perform actions, like generating keys or signing messages, without any single party holding the entire secret. VSS ensures the reliable distribution of key shares among the participants, guaranteeing that even if some shareholders are malicious, the honest parties hold consistent and verifiable parts of the secret key. 

Related works in this domain concentrate on developing efficient and secure threshold signature schemes, threshold encryption schemes, and distributed key generation protocols that rely on VSS as a fundamental building block. 
Li et al.\cite{DKG-VSS} proposed a distributed key generation protocol based on VSS, which reduces the communication overhead during the key generation process. Similarly, TLARDA\cite{TLARDA-audit} introduced a remote data integrity auditing framework in decentralized environments by leveraging two VSS schemes, achieving a significant reduction in communication complexity from $\mathcal{O}(n)$ to $\mathcal{O}(1)$. Ascina~\cite{ascina} utilized VSS to offload tag computation tasks in industrial cloud storage systems, effectively lowering computational overhead.
While these studies have made strides in addressing challenges such as reducing communication and computation costs, ensuring robustness against malicious participants, and enhancing the efficiency of the VSS phase, they still rely heavily on the verification of shares to maintain the consistency of secrets. As a result, there remains significant potential for further improving the overall efficiency of these approaches.

BFT-based VSS is instrumental in achieving consensus and ensuring the integrity of distributed state machines despite the presence of faulty or malicious nodes. 
BFT protocols aim to replicate a service across multiple nodes, such that the system as a whole can tolerate a certain number of Byzantine faults. 
VSS is often employed in the initial stages of these protocols to reliably distribute critical information, such as configuration parameters or initial states, among the participants in a verifiable manner. 

Related research explores how VSS can be adapted and optimized for use in various BFT consensus algorithms, addressing challenges like maintaining consistency in the presence of malicious dealers, ensuring validity, privacy and liveness properties, and minimizing the performance impact of the VSS component on the overall system's efficiency and fault tolerance capabilities.
Although a BFT-based VSS scheme~\cite{basuEfficientVerifiableSecret2019} was recently proposed to force dealers to broadcast the commitments before shares onto a BFT system for maintaining the consistency of the shares. This scheme was still hard to resist against model poisoning attacks based on our exploration that will be shown in Section V.

\section{Preliminaries}
This section briefly recalls verifiable secret sharing, Byzantine Fault Tolerant systems, and VSS-based distributed privacy-preserving machine learning algorithms. All of the algorithms will be used when constructing the SCuMPA and EByFTVeS schemes later. More concretely, the first two algorithms will be used when explaining how malicious dealers can generate secret shares, while the last algorithm will be used to show the serious consequence led by the SCuMPA algorithm, and to design new defense mechanism to deal with the SCuMPA scheme.

Here, the key notations that will be used in the follows are summarized in Table~\ref{tab:notations}.
\begin{table}[H]
  \centering
  \caption{Key Notations}
  \label{tab:notations}
  \renewcommand{\arraystretch}{1.1}
  \begin{tabular}{cl}
    \hline
    Notation & Description \\
    \hline
    $n$       & The total number of participants    \\
    $P_i$       & The participant $i$ \\
    $q, g$ & $g$ is a generator of a cyclic group $\mathbb{Z}_q$ \\
    $p$       & A large prime number that subjects to $q| p-1$\\
    $th$       & The threshold of recovering secret \\
    $M_{i}$   & The machine learning model of the participant $i$ \\
    $T$   & Maximum training round of DPML models\\
    $D_{i}$   & The dataset of the participant $i$  \\
    $Err, \theta$   & The training error and the threshold of model convergence\\
     $\theta_cos$   & The threshold of cosine similarity check\\
    $pk_i, sk_i$    & Public and private key pair of participant $i$  \\
    $\lambda$       & Security Parameter             \\
    $t$         & The current round in training dataset \\
     $w_t^i$     & The model parameter in the round $t$ of participant $i$ \\
    $s^i_j$     & The share of the secret of dealer $i$\\
                & The share is prepared for participant $j$ \\
    $c^i_j$     & The commitment that corresponds to the $j^{th}$ coefficient \\
    & of the polynomial created by dealer $i$\\
    $f$     & The maximum of Byzantine fault (compromised)  participants\\
    $s^i$     & The secret $s$ of participant $i$. $s^i$ equals $w_t^i$ at round $t$.\\
    &  We use this symbol to follow the convention of VSS.\\
    $z_t$ & The number of the reconstructed secrets at round $t$\\
    $\delta$ & The change rate of the model parameters of malicious dealers\\
       $[s^i]$     & The received share of secret $s$ generated by the participant $i$\\
       $[s_i]$     & The aggregated share of secret $s$\\
       & The aggregated share is recovered by participant $i$\\
   $v$ &The current view number\\
   $\delta$ &The hash value of a proposal\\
   $sq$ &The sequence number assigned to the proposal\\
    $u, v$ &The vectors of model parameters of participants\\
    \hline
  \end{tabular}
\end{table}

\subsection{VSS}
VSS \cite{shamir-secret-sharing} comprises of two kinds, the Shamir VSS and additive VSS. In the following sections, we mainly use Shamir VSS as examples, though our scheme is also valid for additive VSS. A typical Shamir secret sharing has two parameters, usually named $f$ and $n$, where $f$ denotes the minimal size of a subset required to recover secret, and $n$ presents the total number of participants. When $f=n$, the function of Shamir VSS can be viewed same as that of additive VSS.

More precisely, Shamir VSS consists of probabilistic polynomial-time algorithms as follows:
\begin{itemize}
\item {Setup$ \rightarrow (p, q, g)$}: Generating a cyclic group $\mathbb{Z}_p$, a subgroup $\mathbb{Z}_q$ of $\mathbb{Z}_p$ , and the generator $g$ of $\mathbb{Z}_q$, where $p$ is a large prime and subject to $q | p - 1$.
	
\item {Share$(s) \rightarrow (\{s_1, \cdots, s_n\}, \{c_0, \cdots, c_{th-1} \})$}: For a secret $s$ in group $\mathbb{Z}_q$ of integers modulo $q$, this algorithm chooses $a_0(s), \cdots, a_{th-1} \leftarrow \mathbb{Z}_q$ randomly and defines a polynomials function: $f(x) = s + \sum_{i=1}^{th-1}a_ix^{i} \mod q$. Additionally, to ensure consistency of shares distributed among participants, this algorithm will generate commitments of shares $c_0=g^s \mod p, c_1 = g^{a_1} \mod p, \cdots, c_{th-1}=g^{a_{th-1}} \mod p $. 
These shares will be sent to the participants.
	
\item {Rect$(s_1, \cdots, s_{th}) \rightarrow s$}: For the shares $s_1, \cdots, s_{th}$, this algorithm recovers the secret $s$ by reconstructing the polynomial below: 
	$$s=\sum_{i}^{th}f(x_i)\frac{\prod_{j=1,j \neq i}^{th} (0-x_j)}{\prod_{j=1,j \neq i}^{th}(x_i - x_j)}$$.
	
 \item {Vrfy$(s_i, \{c_0, \cdots, c_{th-1}\}) \rightarrow true/false)$}: The participant $i$ can verify the received share through checking $g^{s_i} \stackrel{?}{=} c_0 \cdot c_1^i \dots c_k^{i^{th-1}}$. If the result is $true$, the received share is valid.
\end{itemize}

\subsection{Byzantine Fault Tolerance (BFT) System}
\label{sec:pbft}
Byzantine Fault Tolerance (BFT) addresses the classic Byzantine Generals Problem \cite{lamport2019byzantine}, where a group of generals must reach consensus on a battle plan despite the presence of traitors who may send conflicting messages. In BFT systems, this translates to achieving agreement among participants when some might be faulty or malicious, potentially sending inconsistent information to different parts of the network.

Classical BFT systems are categorized based on their network assumptions: synchronous \cite{chan2018pili,abraham2020sync}, asynchronous \cite{duan2023fin,gao2022dumbo}, and partially synchronous. Partially synchronous \cite{pbft,liu2025group,hca} models offer the most practical balance between theoretical guarantees and real-world applicability. We select the Practical Byzantine Fault Tolerance  (PBFT) \cite{pbft} algorithm in this model as the foundation for the design of our consensus-based local gradient exchange algorithm, owing to its fundamental and widely applicable nature. PBFT follows a leader-based consensus paradigm \cite{liu2025abse} and can tolerate up to $f$ Byzantine faults in a system with at least 3$f$+1 participants, requiring a quorum of at least $2f+1$ of the participants to agree.

PBFT proceeds in rounds, with each round consisting of three phases. In the Pre-Prepare phase, a designated leader packages client requests into a proposal and assigns it a sequence number, then broadcasts it via Pre-Prepare messages. During the Prepare phase, each participant that receives a valid Pre-Prepare message broadcasts a corresponding Prepare message to all other participants. Once a participant collects a quorum of matching Prepare messages (including its own), it enters the Commit phase by broadcasting a Commit message. When a participant receives a quorum of matching Commit messages, it executes the proposed operations and responses to client. For any request, the client waits for responses from $f+1$ replicas to finalize it.
To handle leader failures, PBFT also incorporates a View-change mechanism. If participants suspect the leader is faulty (typically due to a timeout), they initiate a view change by broadcasting View-change messages. When a quorum of these messages are collected, a new leader is elected, and the system transitions to a new view where consensus continue with the new leader.

\subsection{The VSS-based Distributed Privacy-preserving Machine Learning}
Distributed privacy-preserving machine learning collaborates multi-participants to train a machine learning model together. It is usually used when the data of each participants are not sufficient enough to make the model convergence. In order to preserve the privacy of the trained parameters, VSS is usually used when participants aggregate all the local trained parameters from her own and others~\cite{ssfl-jsac}. The specification of the VSS-based distributed privacy-preserving machine learning algorithm is listed as below.

\begin{algorithm}[!ht]
    \renewcommand{\algorithmicrequire}{\textbf{Input:}}
	\renewcommand{\algorithmicensure}{\textbf{Output:}}
	\caption{The VSS-based Distributed Privacy-preserving Machine Learning Algorithm}\label{algo-dpml}
    \begin{algorithmic}[1]
    \REQUIRE $Share, Rect, Vrfy, \lambda$, Agg
    \ENSURE $w^i_{t+1}$

    \STATE \textbf{func} SPD($\lambda$)
    \STATE \hspace{\algorithmicindent} Initialize the necessary system parameters
    \STATE \hspace{\algorithmicindent} including $n$, $p$, $th$, $M_i$, $T$, $D_i$, $Err, \theta$.

    \STATE \textbf{func} MPSF($M_i, D_i, t$)
    \STATE \hspace{\algorithmicindent} if ($t == 1$) $w^i_{t-1} = random()$, $z_t=0$
    \STATE \hspace{\algorithmicindent} $w^i_t$ = $M_i(w^i_{t-1}, D_i)$
    \STATE \hspace{\algorithmicindent} $(\{s^i_1, \cdots, s^i_n\}, \{c^i_0, \cdots, c^i_{th-1} \})$=Share$(w^i_t)$
    \STATE \hspace{\algorithmicindent} broadcast $\langle$ $i$, $c^i_0$, $\cdots$, $c^i_{th-1}$$\rangle$
    $\langle$ $i, s^i_1$ $\rangle$, $\cdots$, $\langle$ $i, s^i_n$ $\rangle$ 
    \STATE \hspace{\algorithmicindent} onto the BFT System

     \STATE \textbf{func} MPV($i$, $s^i_1$, $\cdots$, $s^i_n$, $c^i_0$, $\cdots$, $c^i_{th-1}$)
   
      \STATE \hspace{\algorithmicindent} receives $i$, $s^i_1$, $\cdots$, $s^i_n$, $c^i_0$, $\cdots$, $c^i_{th-1}$ from BFT system
      \STATE \hspace{\algorithmicindent} for all ($i,j$) in the commitment
      \STATE \hspace{\algorithmicindent}\hspace{\algorithmicindent} if (Vrfy$(s^i_j$, $\{c^m_0, \cdots, c^m_{z_t-1}\} == true$)
        \STATE \hspace{\algorithmicindent}\hspace{\algorithmicindent} \hspace{\algorithmicindent} save the share $s^i_j$
        \STATE\hspace{\algorithmicindent} \hspace{\algorithmicindent} \hspace{\algorithmicindent}$z_t=z_t+1$
     \STATE \hspace{\algorithmicindent} if ($z_t \geq th$)
     \STATE \hspace{\algorithmicindent}\hspace{\algorithmicindent} compute $s^{i}$ = Rect$(\{s^i_0, \cdots, s^i_{th-1}\})$ 
       \STATE \textbf{func} MPA$(\{s^j | j \in 1, \cdots, n\})$
    \STATE\hspace{\algorithmicindent} 
    if ($|s^j| \geq n-f$)
    \STATE \hspace{\algorithmicindent}\hspace{\algorithmicindent} $w^i_{t+1}$ = Agg$(\{s^j | j \in 1, \cdots, n\})$
        
       \STATE \textbf{func} MPR($w^i_{t+1}$)
	\STATE \hspace{\algorithmicindent} if (Err$(w^i_{t+1}) \leq \theta$ $\OR$ $t == T$)
		     \STATE \hspace{\algorithmicindent}\hspace{\algorithmicindent} return $w^i_{t+1}$
    \end{algorithmic}
\end{algorithm}

The main steps of the Algorithm~\ref{algo-dpml} are introduced as below.

\textbf{Step 1: System Parameter Determination (SPD).} All the participants determine seven kinds of system parameters together. These parameters contain 1) the total number of participants $n$, 2) a large prime $p$, 3) a threshold $th$ for recovering secrets, 4) machine learning models $\{M_i, i \in \{1,...,n\}\}$ with maximum training round $T$ on dataset $\{D_i, i \in \{1,...,n\}\}$, 5) the training error $Err$ and the threshold of the error of model convergence $\theta$, and 6) the aggregation rule Agg of parameters. Here, a classical aggregation rule is to compute the average of all the participant parameters. 7) $\lambda$, the security parameter.

\textbf{Step 2: Model Parameter Sharing and Forwarding (MPSF).} In round $t, t\in\{1,...,N\}$, every participant $p_i$ trains her own machine learning model, and computes her parameter $w^i_t$ based on their sensitive data $D_i$ and random initial gradients $w^i_0$.
After completing model training, every dealer separates her own $w^i_t$ into $n$ shares and broadcasts the shares along with the commitments to every participant.

\textbf{Step 3: Model Parameter Verification (MPV).} 
After every participant receives the commitment $c^i_0$, $\cdots$, $c^i_{th-1}$ and a share $s^i_j$, she runs the function of Vrfy to check the validity of the share. If the result is true, she save the share and increase the number of the received-and-valid shares.

\textbf{Step 4: Model Parameter Aggregation (MPA).} 
After every participant receives $n-f$ secrets from $\{s^1,...,s^{n}\}$, she computes the aggregation due to the predetermined aggregation rule. The aggregated result is typically the average of all the received shares in round $t$. 

\textbf{Step 5: Model Parameter Reconstructing (MPR).} 
Every participant checks whether the machine learning model can converge based on the aggregated parameter or reaches the maximal training round $T$. If so, output the parameter and stop. Otherwise, go to a new round $t+1$.

\begin{figure}[!t]
    \includegraphics[width=4.5in]{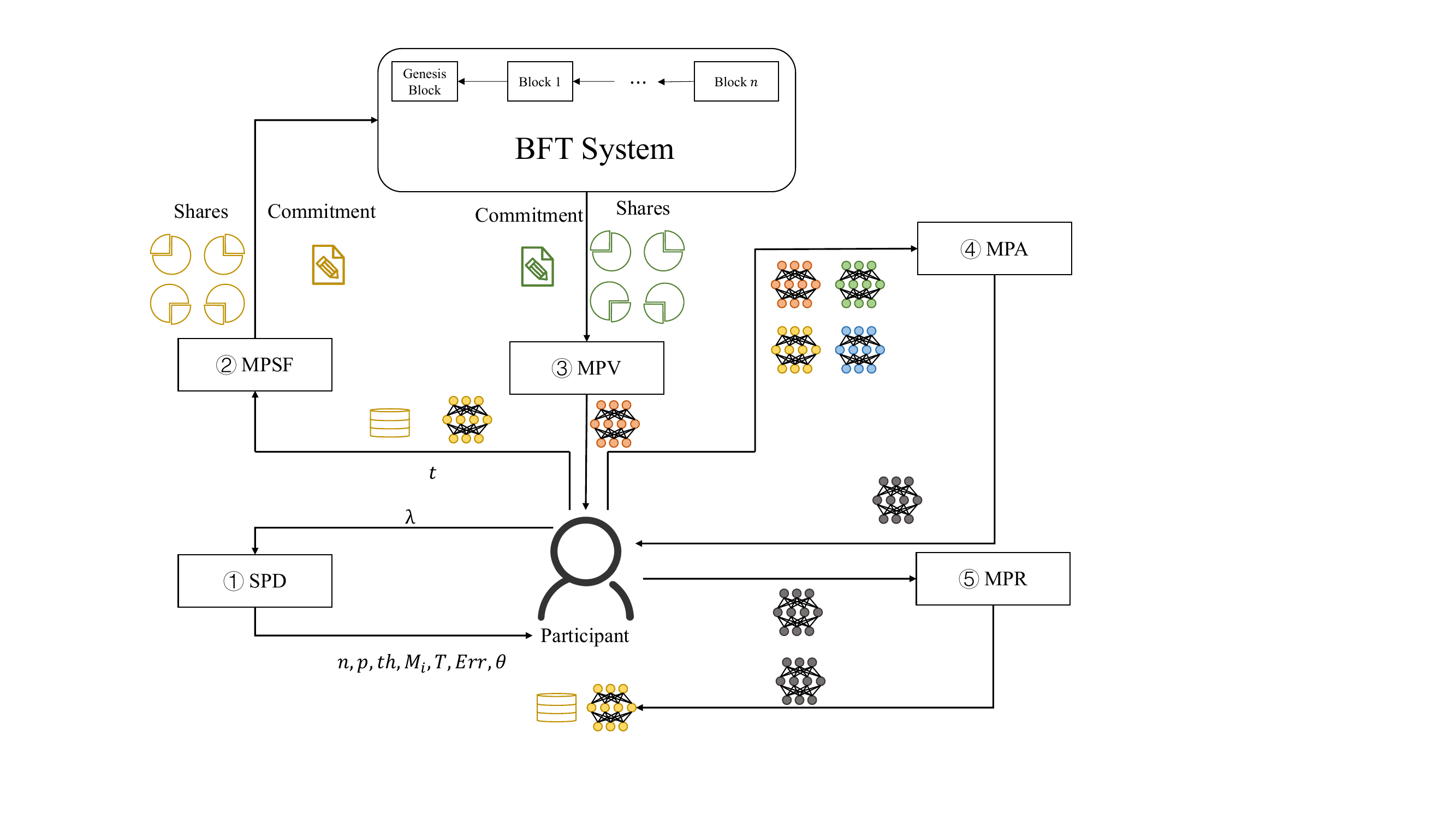}
    \caption{The Flowchart of the BFT-based VSS in DPML}\label{fig-dpml}
\end{figure}

\section{Models and Goals}
\subsection{System Model}
A typical Byzantine fault tolerant-based verifiable secret sharing scheme in distributed privacy-preserving machine learning is assumed to consist of $n$ participants. All of the participants communicate with each other to forward shares of secret. When sharing the shares, the participants are common to be named as dealers. After that, participants compute the average of secrets of dealers according to the received shares afterwards. The flowchart of the entire system is as shown in Figure~\ref{fig-dpml}. The fundamental system parameters and functions can follow the input of Algorithm~\ref{algo-dpml}. 

Here, we make a partial synchronous network assumption in the system: there exists an unknown Global Stabilization Time (GST) after which all messages between honest participants are delivered within a bounded delay $\Delta$. The network connection is established from point to point among participants, which is reliable and authenticated by the public identity. Besides, in a period of time, generally called a view, there is a participant designated as the leader to drive the protocol for one round or more.

\subsection{Threat Model}
We assume up to $f$ participants who can be compromised, where $f$ does not exceed $\frac{n-1}{3}$ of all participants. These compromised participants are Byzantine faulty with arbitrary behavior, meaning they can deviate from the protocol in any way, including forwarding different shares to different participants or remaining silent as they play role of dealers. However, they cannot alter messages after transmission or break down underlying cryptographic primitives in the probabilistic polynomial time of a certain security parameter $\lambda$.

Apart from the compromised participants, an adversary can also eavesdrop on the forwarded messages, and control the arrival time of those messages, even delete the messages when necessary, for assisting the compromised participants to launch model poisoning attacks. That is, colluding is permit for compromised participants and eavesdroppers.

\subsection{System Goal}
Our system aims to achieve the following properties:
\begin{itemize}[]
\item \textbf{Validity,} which ensures that all honest participants agree on the sequence number of requests that commit locally.

\item \textbf{Liveness,} which guarantees that every training round can continue to move forward even in the presence of faulty participants.
	
\item \textbf{Consistency,} which guarantees that the computed results of honest participants remain consistent before and after each training round.
	
\item \textbf{Privacy,} which ensures that all secrets of dealers can be protected during in the steps of share forwarding and share aggregating among authorized participants.
\end{itemize}

\section{Adaptive Share Delay Provision}

In this section, we explore an Adaptive Share Delay Provision (ASDP) strategy. This strategy is used for malicious dealers to provide bad model parameters for launching model poisoning attack. Here, the model parameters indicate gradients for various artificial neural networks. The divergence of the distributed privacy-preserving machine learning can be delayed inevitably. We also show the theoretical analysis to prove the feasibility of the proposed ASDP strategy. 

Since each participant has to independently aggregate $th$ shares by running the function of MPA in algorithm~\ref{algo-dpml} for training the model, malicious dealers can meticulously elaborate their model parameters to resist against the common model poisoning prevention mechanisms, like cosine similarity-based \cite{fltrust} or $L$ regularization-based \cite{krum} gradient aggregation. 

Here, we take an example to illustrate the procedure of the ASDP strategy. Assume that there are four participants, and the first participant is assumed to be a malicious dealer, while others are viewed as honest. The core idea of the malicious dealer is to try to provide a bad system parameter in each round to delay the model divergence as long as possible. The provided gradients can not only make the direction of the targeted gradient as far as the correct direction to a certain victim participant, but also pass through the model poisoning prevention check of this participant. More precisely, a malicious dealer can recover the  gradient $w^i$ at first by using the shares of $P_i$ obtained from the BFT system. Next, the malicious dealer chooses an orthogonal gradient of $w$ at the start, and then keep slightly adjusting the angle till $cos(w, w') \leq \theta_{cos}$, or $L(w, w') \leq \theta_{L}$, where $cos$ and $L$ are standard functions to compute cosine similarity, and $L$ regularization of the aggregated gradient and the correct gradient, as shown in Figure~\ref{fig:ASDP}. Here, $\theta_{cos}$ and $\theta_{L}$ are predefined super-parameters of the system. The concrete procedure of the ASDP strategy is shown in Algorithm~\ref{algo-adlgg}.

\begin{algorithm}[!ht]
    \renewcommand{\algorithmicrequire}{\textbf{Input:}}
	\renewcommand{\algorithmicensure}{\textbf{Output:}}
	\caption{The procedure of the Adaptive Share Delay Provision Strategy}\label{algo-adlgg}
    \begin{algorithmic}[1]
		\REQUIRE $w^i$, $\theta_{cos}$, $\delta$; 
		\ENSURE $w'^i$;
            \STATE target\_norm=\textbf{norm}($w^i$)
            \STATE target\_abs = \textbf{abs}($w^i$)
            \STATE target\_sign = \textbf{sign}($w^i$)
            \STATE indices = \textbf{sort}(target\_abs)
            \STATE $w'^i$=$\textbf{zeros\_like}(w^i)$
            \STATE norm\_squared=0
            \STATE indicator=0
            \FORALL {each $idx \in indices$}
			\STATE sign = target\_sign[idx]
			\STATE $w'^i[idx$] = sign
			\STATE indicator += $w^i$[idx]  $\times$ $w'^i$[idx]
			\STATE norm\_squared += $\delta$
                \STATE norm = $ \sqrt{\text{norm\_squared}} $
                \STATE cos = $\frac{\text{indicator}}{\text{target\_norm} \times \text{norm}}$
			\IF {cos $\le \theta_{cos}$} 
			\STATE break
			\ENDIF
		\ENDFOR
		\STATE norm = $\sqrt{\text{norm\_squared}}$
		\STATE factor = $\frac{\text{target\_norm}(w_j)}{\text{norm}}$
        \STATE $ w'^i = w'^i \times$ factor
		\RETURN $w'^i$
    \end{algorithmic}
\end{algorithm}

\subsection{Theoretical Analysis}
Informally, the ASDP strategy is valid because most of the existing distributed privacy-preserving machine learning schemes do not force the participants to only use those gradients of which the commitment is provided in a certain time period. Thus, a malicious user can aggregate the gradients at the start, and provide the commitment with the share at the very last moment. Since the direction of $w$ and $w'$ is as far as possible, the training error must become large enough to increase the training rounds, thereby causing the training difficult to diverge. The specific formal analysis is presented in the Appendix A (supplementary material).

\section{The ASDP-based Customized Model Poisoning Attack}
In this section, we launch an ASDP-based Customized Model Poisoning Attack (ACuMPA). Malicious dealers can compute a customized gradident using ASDP strategy for a certain victim participant. We also prove the validity of the ACuMPA scheme.

\subsection{The ACuMPA Scheme}
\begin{figure}[!b]
    \centering
    \includegraphics[width=3.5in]{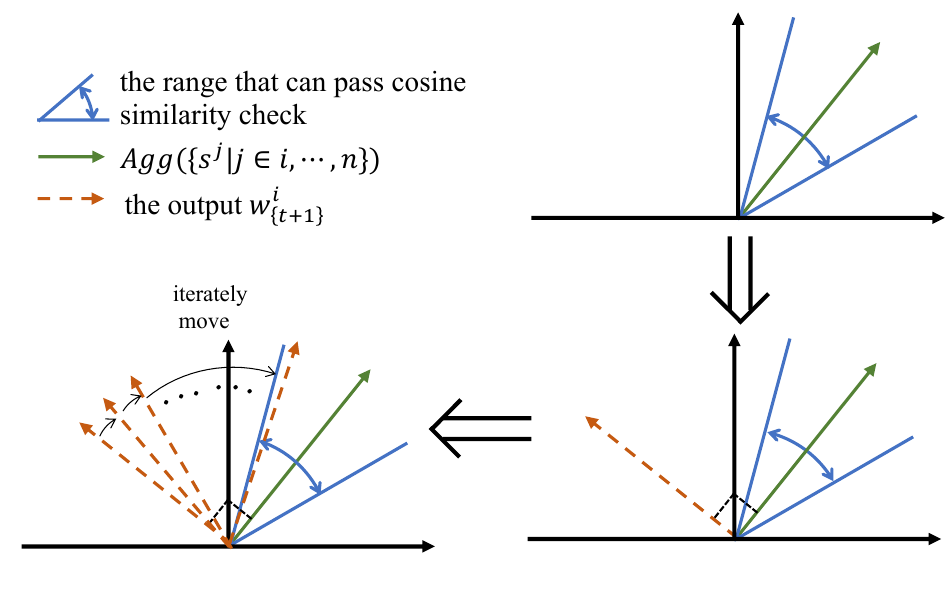}%
    
    \caption{The Core Procedure of the ACuMPA Scheme}\label{fig:ASDP}
\end{figure}

Since the ASDP strategy can enable malicious dealers to generate bad model parameters to slow down the speed of model convergence, we can apply this strategy to launch model poisoning attack. The specification of the ACuMPA scheme is presented in Algorithm~\ref{algo-miaf-framework}. Specifically, the algorithm initially identifies for orthogonal vectors for parameters received, then diverges the results by a very small angle in the other direction as shown in Figure~\ref{fig:ASDP}.

\begin{algorithm}[!ht]
    \renewcommand{\algorithmicrequire}{\textbf{Input:}}
	\caption{The procedure of the ASCuMPA Scheme}\label{algo-miaf-framework}
        \begin{algorithmic}[1] 
        \REQUIRE $n$ participants $p_1, p_2, \cdots, p_n$ with local training datasets $D=\{D_1, D_2, \cdots, D_n\}$, training epoch $T$;
	\FORALL {each $t \in [1,T]$}
			\FORALL {each $i \in [1,n]$}
			\STATE Participant trains a model $w^i_t$ based on dataset $D_i$.
			\IF {$p_i$ is honest participant}
			\FORALL {$p_j$}
				\STATE participant $p_i$ send the shares of $s^i_t$ and the corresponding commitments to $p_j$.
			    \STATE Participant receives parameters $\{s^1_t,\cdots, s^n_t\}$
                    \FORALL {$p_j \in \mathcal{N}_{i}$}
                        \STATE Participant $p_i$ verifies the validity of $s^j_t$. 
                    \ENDFOR
			\ENDFOR
                \ELSE
                    \STATE Malicious participants reconstruct and aggregate the average of model parameters using Algorithm~\ref{algo-dpml}
                    \STATE Malicious participants generate bad gradients using Algorithm~\ref{algo-adlgg}
			\ENDIF
                \ENDFOR
			\STATE Participants aggregate and update model $w^t$ locally.
	\ENDFOR
    \end{algorithmic}
\end{algorithm}

\subsection{Theoretical Analysis}


\begin{theorem}[Existence of Common Substitute Vector]\label{thm:common-sub}
Let $\Delta=\tfrac{1}{N}\sum_{i=1}^{N}v_i$ be the global gradient and let $\mathcal{S}$ denote the index set of the top-$k$ coordinates w.r.t.\ $|\Delta_i|$.  
For any client $i$, define the normal similarity lower bound  
\[
\tau_0=\frac{\|v_i\|_1}{\|v_i\|\sqrt{d}} .
\]
If $k\ge C(\alpha)\log d$ for a constant $C(\alpha)$ depending only on the Dirichlet concentration parameter $\alpha$, then there exists a vector  
$u\in\{-1,0,1\}^d$ with $\mathrm{supp}(u)=\mathcal{S}$ such that
\[
\cos(v_i,u)\;\ge\;\tau_0
\quad\text{for at least }\;\Omega(N)\text{ clients.}
\]
\end{theorem}

\begin{proof}
Let $u=\operatorname{sign}(\Delta_{\mathcal{S}})$ be the sparse sign vector of the global gradient.  
For any client $i$, by the Cauchy–Schwarz inequality,
\[
\langle v_i,u\rangle
  =\langle\Delta,u\rangle+\langle v_i-\Delta,u\rangle
  \;\ge\;
  \|\Delta_{\mathcal{S}}\|_{1}-\|v_i-\Delta\|\,\|u\|.
\]
Since the normal similarity bound yields $\|v_i-\Delta\|\le\delta(\alpha)$ and $\|u\|=\sqrt{k}$, we have
\[
\cos(v_i,u)\;\ge\;
\frac{\|\Delta_{\mathcal{S}}\|_{1}-\delta(\alpha)\sqrt{k}}{\|v_i\|\sqrt{k}} .
\]
Dirichlet gradient concentration implies $\|\Delta_{\mathcal{S}}\|_{1}\ge(1-\varepsilon)\|\Delta\|_{1}$ whenever $k\ge C(\alpha)\log d$.  
Combining this with Lemma 1’s maximum-cosine bound,
\[
\cos(v_i,u)\;\ge\;
\underbrace{\frac{\|\Delta\|_1}{\|\Delta\|\sqrt{d}}}_{s_{\max}(\Delta)}
          \cdot\frac{1-\varepsilon}{\sqrt{k/d}}
          -\frac{\delta(\alpha)}{\|v_i\|}.
\]
Setting $k=\Theta(\log d)$ and assuming 
$\delta(\alpha)\le\|v_i\|\,s_{\max}(\Delta)/2$
make the right-hand side exceed $\tau_0$.  
A standard covering-number argument then shows that at least $\Omega(N)$ clients satisfy $\cos(v_i,u)\ge\tau_0$ simultaneously.
\end{proof}

\section{The [E]fficient [By]zantine [F]ault [T]orlerant-based [Ve]rifiable [S]ecret-sharing (EByFTVeS) Scheme}
In this section, we apply a four-phase consensus algorithm based on Practical Byzantine Fault Tolerance (PBFT) \cite{pbft} to propose the EByFTVeS scheme against model poisoning attack. We then analyze the validity, liveness, consistency, and privacy properties of the EByFTVeS scheme.

\subsection{The EByFTVeS Algorithm}
Malicious dealers can carefully craft gradients to certain participants based on Algorithm~\ref{algo-miaf-framework}. To prevent this attack, we force all the dealers provide the shares with the  commitments in a certain time period. Later, only the gradients from these shares are used to aggregate.

The concrete procedure of the EByFTVeS Algorithm is illustrated in Algorithm \ref{alg:ebyftves}, with our modified PBFT-based consensus Algorithm (Algorithm \ref{alg:consensus}) serving as the main part. We replace the traditional "broadcast" and "receiving" operations in Algorithm \ref{alg:ebyftves} with "broadcast\_update" and "receiving\_update" functions from our consensus Algorithm. Specifically, when a participant needs to share their secret shares (line 8 in Algorithm 6), verification results (line 14), aggregated shares (line 21), they invoke "broadcast\_update" instead of direct broadcasting, which ensures that all messages are processed through the consensus mechanism. 

PBFT \cite{pbft} was chosen as it is a fundamental and widely applicable BFT protocol. However, we note that this is not the only feasible option, and other BFT protocols \cite{dolphin,hotstuff} could also be used. 
In the normal-case operation of the classical PBFT, participants directly submit requests to a primary participant, a process that is not feasible in the context of the DPML. To address this, we introduce a modification by adding an additional \textbf{Pre-Propose} phase before the conventional three phases of \textbf{Pre-Prepare}, \textbf{Prepare}, and \textbf{Commit}. Note we only sketch the major workflow for normal-case operation here; mechanisms such as checkpointing are excluded on account of their alignment with the original protocol (the array we created should also be considered during garbage collection).

\begin{algorithm}
	\caption{The Consensus Algorithm}
	\label{alg:consensus}
	\begin{algorithmic}[1]
		\STATE \textbf{Initialization}
		\STATE \hspace{\algorithmicindent}$v$, $f$ \COMMENT{{\color{gray}view number, maximum faulty nodes}}
		\STATE \hspace{\algorithmicindent}$batch\_size$ \{{\color{gray}minimum valid batch size}\}
		\STATE \hspace{\algorithmicindent}$P$ \{{\color{gray}set of all participants}\}
		\STATE \hspace{\algorithmicindent}$pending\_requests$ \{{\color{gray}set of pending requests}\}
		\STATE \hspace{\algorithmicindent}$initial\_proposals$ \{{\color{gray}set of initial proposals}\}
		\STATE \hspace{\algorithmicindent}$initial\_proposals\_raw$
		\STATE \hspace{\algorithmicindent}$props$
		
		\STATE \textbf{as a participant}
		\STATE \textbf{func} broadcast\_update($sq, \textsc{req}\langle req\rangle$) \COMMENT{{\color{gray}$sq$ is the sequence number}}
		\STATE \hspace{\algorithmicindent} broadcast $\langle$\textsc{request}, $v, sq$, $req\rangle$ to $P$
		
		\STATE \textbf{upon} receiving a valid $m$$ =$$ \langle$\textsc{request}$, v,$$ sq$,$req\rangle$ from $p_j$
		\STATE \hspace{\algorithmicindent} $pending\_requests[sq][j] \leftarrow m$
		\STATE \hspace{\algorithmicindent} if $|pending\_requests[sq]| \geq batch\_size$
		\STATE \hspace{\algorithmicindent}\hspace{\algorithmicindent} $props[sq] \leftarrow pending\_requests[sq]$
		\STATE \hspace{\algorithmicindent}\hspace{\algorithmicindent} $pending\_requests[sq]\leftarrow []$
		\STATE \hspace{\algorithmicindent}\hspace{\algorithmicindent} send $\langle$\textsc{pre-propose}, $v, sq$, $props[sq]\rangle$ to primary($v$)
		
		\STATE \textbf{upon} receiving a valid $m=\langle$\textsc{pre-propose}, $v,sq$, $prop_j\rangle$ from $p_j$, $initial\_proposals\_raw[sq][j]\leftarrow m$
		\STATE \hspace{\algorithmicindent} $initial\_proposals[sq][j] \leftarrow prop_j$
		\STATE \hspace{\algorithmicindent} if $|initial\_proposals[sq]| > 2f$
		\STATE \hspace{\algorithmicindent}\hspace{\algorithmicindent} $batch \leftarrow$ aggregate($initial\_proposals[sq]$)
		\STATE \hspace{\algorithmicindent}\hspace{\algorithmicindent} if primary($v$)
		\STATE \hspace{\algorithmicindent}\hspace{\algorithmicindent}\hspace{\algorithmicindent}broadcast $\langle$\textsc{pre-prepare}, $v$, $sq$, $h(batch),$
		\STATE \hspace{\algorithmicindent}$\hspace{\algorithmicindent}\hspace{\algorithmicindent} initial\_proposals\_raw[sq]\rangle$ to $P$
		\STATE \hspace{\algorithmicindent}$\hspace{\algorithmicindent}\hspace{\algorithmicindent}$$initial\_proposals[sq]\leftarrow []$ \STATE \hspace{\algorithmicindent}$\hspace{\algorithmicindent}\hspace{\algorithmicindent}$$initial\_proposals\_raw[sq]\leftarrow []$
		
		\STATE \textbf{upon} receiving a valid $m = \langle$\textsc{pre-prepare}, $v$, $sq$, $h(batch), iRaw\rangle$ from primary($v$), if valid sequence number $sq$ and not already accepted for $v$, $sq$
		\STATE \COMMENT{{\color{gray} the correctness of $batch$ is verified by performing a computation similar to creating $batch$}}
		\STATE \hspace{\algorithmicindent}\hspace{\algorithmicindent} broadcast $\langle$\textsc{prepare}, $v$, $sq$, $h(batch)\rangle$ to $P$
		
		\STATE \textbf{upon} receiving $2f+1$ matching $\langle$\textsc{prepare}, $v$, $sq$, $\delta\rangle$
		\STATE \hspace{\algorithmicindent} $prepared(\delta, v, sq) \leftarrow true$
		\STATE \hspace{\algorithmicindent} broadcast $\langle$\textsc{commit}, $v$, $sq$, $\delta\rangle$ to $P$
		
		\STATE \textbf{upon} receiving $f+1$ matching $\langle$\textsc{commit}, $v$, $sq$, $\delta\rangle$
		\STATE \hspace{\algorithmicindent} broadcast $\langle$\textsc{commit}, $v$, $sq$, $\delta\rangle$ to $P$
		
		\STATE \textbf{upon} receiving $2f+1$ matching $\langle$\textsc{commit}, $v$, $sq$, $\delta\rangle$ where $h(batch) = \delta$
		\STATE \hspace{\algorithmicindent} execute($batch$)
		\STATE \hspace{\algorithmicindent} $committed$-$local(\delta, v, sq) \leftarrow true$
		
		\STATE \textbf{func} aggregate($iProposals$)
		\STATE \hspace{\algorithmicindent} $batch \leftarrow \emptyset$
		\STATE \hspace{\algorithmicindent} for each request $r$
		\STATE \hspace{\algorithmicindent}\hspace{\algorithmicindent} if $r$ appears in $> f$ proposals in $proposals$
		\STATE \hspace{\algorithmicindent}\hspace{\algorithmicindent}\hspace{\algorithmicindent} $batch \leftarrow batch \cup \{r\}$

		\STATE \hspace{\algorithmicindent} return $batch$
		
		\STATE \textbf{func} execute($batch$)
		\STATE \hspace{\algorithmicindent} for each $req \in batch$
		\STATE \hspace{\algorithmicindent}\hspace{\algorithmicindent} receiving\_update$(req)$

	\end{algorithmic}
\end{algorithm}

\begin{algorithm}
	\caption{The Procedure of EByFTVeS Algorithm}
	\label{alg:ebyftves}
	\begin{algorithmic}[1]

           \REQUIRE  Share, Rect, Vrfy, $\lambda$, Agg, Enc, Dec
    \ENSURE $w^i_{t+1}$

    \STATE \textbf{func} SPD($\lambda$)
    \STATE \hspace{\algorithmicindent} Initialize the necessary system parameters
    \STATE \hspace{\algorithmicindent} including $n$, $p$, $th$, $M_i$, $T$ , $D_i$, $Err, \theta$, $pk_i, sk_i$.

    \STATE \textbf{func} MPSF($M_i, D_i, t$)
    \STATE \hspace{\algorithmicindent} if ($t == 1$) $w^i_{t-1} = random()$
    \STATE \hspace{\algorithmicindent} $w^i_t$ = $M_i(w^i_{t-1}, D_i)$
    \STATE \hspace{\algorithmicindent} $(\{s^i_1, \cdots, s^i_n\}, \{c^i_0, \cdots, c^i_{th-1} \})$=Share$(w^i_t)$
    \STATE \hspace{\algorithmicindent} {\color{purple}broadcast\_update} $(3t, \textsc{req}\langle$ $i$, Enc$(pk_1, s^i_1)$,
    \STATE \hspace{\algorithmicindent}  $\cdots$, Enc$(pk_n,s^i_n)$, $c^i_1$, $\cdots$, $c^i_{n}$$\rangle)$

     \STATE \textbf{func} MPV($sk_i, m,$ Enc$(pk_i, s^m_i), \{c^m_0, \cdots, c^m_{th-1}\}$)
      \STATE \textbf{upon} {\color{purple}receiving\_update} $m$, Enc$(pk_i, s^m_i)$, $c^m_0, \cdots, c^m_{th-1}$ by the Consensus Algorithm {\color{purple}under sequence number $sq$}
     \STATE \hspace{\algorithmicindent}\hspace{\algorithmicindent} $[s^m]$ = Dec($sk_i$, Enc$(pk_i, s^m_i)$)
      \STATE \hspace{\algorithmicindent}\hspace{\algorithmicindent} if (Vrfy($[s^m]$, $\{c^m_0, \cdots, c^m_{th-1}\}) == true$)
       \STATE \hspace{\algorithmicindent}\hspace{\algorithmicindent}\hspace{\algorithmicindent} {\color{purple}broadcast\_update} $(sq+1, \textsc{req}\langle$$(verified, m, i)$$\rangle)$

       \STATE \textbf{func} MPA$(\{(verified, m, l)\}$)
     \STATE \textbf{upon} {\color{purple}receiving\_update} $th$ verified messages by querying
     \STATE \hspace{\algorithmicindent} $(verified, m, i)$ for all $m, i$ on Consensus Algorithm {\color{purple}} 
     \STATE \hspace{\algorithmicindent} {\color{purple}under sequence number $sq$}
        \STATE \hspace{\algorithmicindent} save the $th$ shares as $[s^1], [s^2],...,[s^{z_t}]$
        \STATE \hspace{\algorithmicindent} $[s_i]$ = Agg$([s^1], [s^2],...,[s^{z_t}])$
        \STATE \hspace{\algorithmicindent} {\color{purple}broadcast\_update} $(sq+1,\textsc{req}\langle$$([s_i], i)$$\rangle)$

       \STATE \textbf{func} MPR($\{([s_j], j)\}$)
        \STATE \textbf{upon} there exists $th - 1$ $\{([s_j], j) | j \neq i\}$ on blockchain
        \STATE \hspace{\algorithmicindent} $w^i_{t+1}$ = Rect$([s_1], [s_2],...,[s_{th}])$
	\STATE \hspace{\algorithmicindent} if (Err$(w^i_{t+1}) \leq \theta$ $\OR$ $t == T$)
		     \STATE \hspace{\algorithmicindent}\hspace{\algorithmicindent} return $w^i_{t+1}$
	\end{algorithmic}
\end{algorithm}

In our consensus Algorithm, all messages are independently initiated by participants and submitted via broadcast\_update. Each participant collects messages independently. Once the number of collected requests (under the same sequence number) exceeds the configurable batch size (set to greater than two-thirds of the total replicas in our implementation, but can be modified depending on $th$), the node packages them into an initial proposal. This proposal is then send to the primary as part of the \textbf{Pre-Propose} phase in a \textbf{Pre-Propose} message.

When the primary receives valid \textbf{Pre-Propose} messages (under the same sequence number) from more than two-thirds of the node, it aggregates the initial proposals contained in these messages into a formal proposal. Message that appear in greater than one-third of the initial proposals are included in the formal proposal. The primary then broadcasts the formal proposal in a \textbf{Pre-Prepare} message. Subsequently, the Algorithm transitions into the three-phase consensus process (\textbf{Pre-Prepare}, \textbf{Prepare}, and \textbf{Commit}), which is similar to the original PBFT protocol (see Section \ref{sec:pbft}).

The view-change procedures remain consistent with the original PBFT protocol (the pseudocode is omitted here to maintain brevity and highlight our modifications). When participants suspect the primary is faulty (typically due to a timeout), they initiate a view change by broadcasting View-change messages. When $2f+1$ valid View-change messages are collected, a new primary is elected, and the system transitions to a new view where consensus can continue. 



\subsection{Theoretical Analysis}
\begin{theorem} [Validity]
\label{the:safety}
All honest participants agree on the sequence number of requests that commit locally.
\end{theorem}

\begin{proof}
We consider both intra-view and inter-view scenarios.
\vspace{1mm}
\noindent Case 1 (Intra-view Validity:): We prove this by contradiction. Suppose honest participants $p_i$ and $p_j$ commit different proposals $prop$ and $prop'$ respectively with the same sequence number $sq$ in view $v$. For participant $p_i$ to commit $prop$, it must have received $2f + 1$ matching Commit messages for some digest $\delta = h(prop)$ (i.e., $committed\text{-}local(\delta,v,sq)=true$). 
Similarly, for participant $p_j$ to commit $prop'$, it must have received $2f + 1$ matching Commit messages for some digest $\delta' = h(prop')$ (i.e., $committed\text{-}local(\delta',v,sq)=true$). 
Since there are at most $f$ Byzantine participants, By the quorum intersection property ($2(2f+1) - n = (4f+2) - (3f+1) = f+1$), there exists at least one honest participant $p^*$ that sent both Commit messages for $\delta$ and $\delta^\prime$.  However, honest participants only send one Commit message per sequence number per view, leading to a contradiction.

\vspace{1mm}

\noindent Case 2 (Inter-view Validity:): For inter-view safety, we must show that if $p_i$ commits $prop$ in view $v_1$ with $sq$, and $p_j$ commits $prop'$ with $sq$ in a later view $v_2 > v_1$, then $prop = prop'$. Since our view-change mechanism is inherited from PBFT, it ensures that: 

\noindent (1) Any request from a previous view that was committed but uncompleted (i.e. whose corresponding training round has not been completed in our scheme) is carried forward to the new view. When a new primary is elected in $v_2$, it must collect $2f+1$ View-change messages to create a New-view message. Since at most $f$ participants are faulty, and any committed request requires responses from at least $f+1$ participants, there must be at least an honest participant who had committed the request in $v_1$ to include this information (together with the corresponding certificate) in the View-change message and must be processed in the new view due to the validity requirements for new-view messages (otherwise honest participants will not accept the message and attempt to initiate a new view change).

\noindent (2) Prior to the start of the normal-case operation in the new view, any uncommitted but prepared requests (also included in the participants' view change messages and must be attached to corresponding Pre-prepare messages, see the original PBFT for details) are redone through the standard three-phase consensus process (i.e. excluding the Pre-propose phase) with the original sequence number and the new view number.

Thus, v
alidity is also preserved across views.
\end{proof}

\begin{theorem} [Liveness] After GST, every training round can continue to move forward even in the presence of up to $f$ faulty participants.
\label{the:liveness}
\end{theorem}

\begin{proof}
   We prove liveness by showing that every request broadcast by an honest participant will eventually be committed by all honest participants. Let $t_{GST}$ be the Global Stabilization Time after which all messages between honest participants are delivered within time $\Delta$. Consider a request $req$ broadcast by an honest participant $p_i$ at time $t \geq t_{GST}$.

   \noindent Case 1 (Honest Primary): If current primary is honest, then:
   \begin{enumerate}
       \item All honest participants receive the Requst message within time $\Delta$ and add it to their pending requests.
       \item When an honest participant's pending requests for sequence number $sq$ reach the batch size threshold (which is set to more than $2f$), it sends a Pre-propose message to the primary.
       \item Since there are $2f + 1$ honest participants and at most $f$ Byzantine participants, at least $2f + 1 > 2f$ honest participants will eventually send Pre-propose messages containing $req$.
       \item The honest primary will receive more than $2f$ Pre-propose messages within time $\Delta$ and will include $req$ in the aggregated batch (since $req$ appears in at least $2f+1-f=f$ proposals).
       \item The primary broadcasts the Pre-prepare message, and the standard PBFT three-phase algorithm ensures commitment within bounded time.
   \end{enumerate}

    \noindent Case 2 (Byzantine Primary): If the current primary is Byzantine, it may send invalid or conflicting Pre-prepare messages or may delay them indefinitely. However:
    \begin{enumerate}
        \item Honest participants implement timeout mechanisms. If more than $f$ of them do not receive a valid Pre-prepare message or cannot collect $2f+1$ distinct Prepare/Commit messages within the timeout period, a view change is initiated eventually.
        \item The view change protocol (identical to PBFT) ensures that after $t_{GST}$, a new view with an honest primary will be established within bounded time.
        \item The new honest primary will process pending requests as described in Case 1.
    \end{enumerate}

\vspace{1mm} Since at most $f$ out of $3f + 1$ participants are Byzantine, there are at least $2f + 1$ honest participants. In the worst case, we may need to go through at most $f$ views with Byzantine primaries before reaching an honest primary. Each view change completes within bounded time after $t_{GST}$, ensuring eventual progress.
\end{proof}

\begin{theorem} [Consistency]
 The EByFTVeS scheme is secure against the ACuMPA attack and maintains consistency of computed results across all honest participants.
\end{theorem}

\begin{proof}
The ACuMPA attack relies on the ability of malicious dealers to provide inconsistent shares to different participants without detection. However, the EByFTVeS scheme prevents this through its consensus-based message consistency mechanism. In the traditional VSS-based DPML (Algorithm 1), a malicious dealer could send different shares to different participants, exploiting the lack of cross-validation among participants. The ACuMPA attack leverages this by using the ASDP strategy to craft customized malicious gradients for specific victims. The EByFTVeS scheme counters this attack vector through the following mechanisms:

\begin{enumerate}
    \item All share distribution goes through the consensus algorithm (satisfying Safety and Liveness, as outlined in Theorems \ref{the:safety}-\ref{the:liveness}). When a dealer broadcasts shares via "broadcast\_update" in a training round $t$, the consensus mechanism ensures that all honest participants receive (via "receiving\_update") the same set of shares of $t$.
    \item Shares are encrypted with recipients' public keys before consensus, preventing unauthorized modification while maintaining privacy. The verification process ensures that only valid shares are accepted.
    \item The verification results are also broadcast through consensus, ensuring that all participants have a consistent view of which shares are valid.
    \item The final aggregation step also uses consensus, preventing malicious participants from providing different aggregation results to different participants.
\end{enumerate}

Since the consensus algorithm guarantees that all honest participants receive identical messages (by Theorem \ref{the:safety}), malicious dealers cannot execute the differential share distribution required for the ACuMPA attack.
\end{proof}

\begin{theorem} [Privacy]
If the VSS-based scheme preserves privacy of secrets of dealers, then the EByFTVeS scheme can preserve privacy of the secrets.
\end{theorem}

\begin{proof}
Assume that there is an adversary $\mathcal{A}$ who runs the EByFTVeS scheme and attempts to can violate the privacy of secrets of dealers with the probability of $\epsilon(\lambda)$. A simulator $\mathcal{S}$ can call $\mathcal{A}$ to attack the VSS scheme as below.

\begin{enumerate}
    \item $\mathcal{S}$ first randomly picks up a dealer $i$ and tries to violate the privacy of the secret of this dealer.

\item When $\mathcal{A}$ requests to communicate with any participant $j$, $\mathcal{S}$ checks whether this participant exists or not. If exists, $\mathcal{S}$ send the request to the participant. Otherwise, if $j \neq i$, $\mathcal{S}$ create this participant and then send the request to the participant. If $j = i$, $\mathcal{S}$ call the VSS scheme to create the participant, and then send the corresponding request in the way of the VSS scheme to the participant.

\item When When $\mathcal{A}$ picks up a dealer $j$ to challenge the privacy of secret. If $j \neq i$, $\mathcal{S}$ aborts. Otherwise, output whatever $\mathcal{A}$ outputs.
\end{enumerate}

The EByFTVeS scheme has only one difference when comparing with the traditional VSS-based scheme. All of the requests are encrypted and then sent to the BFT system. But when the corresponding participant obtains the request from the system, decrypts the request, and then sends the plaintext as the corresponding request which has the same way in the VSS scheme, this difference does not impact when violating the privacy of the secret. The remainder is whether $j$ equals $i$. Since $i$ is randomly selected and the probability of privacy violation is $\epsilon(\lambda)$, the probability of privacy violation becomes $\epsilon(\lambda)/N$. If the privacy violation for the EByFTVeS scheme is not negligible, it is also not negligible for the VSS-based scheme.
\end{proof}

\section{Performance Analysis}

In this section, we setup a series of experiments to quantitatively check the performance of the EByFTVeS scheme.



\subsection{Evaluation on DML}
\subsubsection{Experimental Settings}
We introduce the experimental settings of evaluations on distributed machine learning.


{\bf Settings.} We implement a prototype of the proposed schemes using PyTorch and Python 3.8. All experiments are conducted on a server running Ubuntu 22.04, equipped with a 16-core Intel(R) Xeon(R) Gold 6430 CPU, 120 GB of RAM, and an NVIDIA GeForce RTX 4090 GPU.

{\bf Datasets and Models.} We evaluate the performance of the proposed ACuMPA scheme on three widely used visual recognition benchmark datasets: MNIST~\cite{lecun1998mnist}, Fashion-MNIST~\cite{xiao2017fashion}, and CIFAR-10~\cite{krizhevsky2009learning}. 

MNIST is a classic dataset of handwritten digit images, comprising 60,000 training samples and 10,000 testing samples distributed evenly across 10 classes (digits 0–9). It serves as a fundamental benchmark for image classification tasks involving grayscale images of size $28 \times 28$ pixels. 

Fashion-MNIST is a more challenging dataset intended as a direct drop-in replacement for MNIST. Released by Zalando Research, it consists of 70,000 grayscale images of fashion products categorized into 10 classes such as t-shirts, trousers, and ankle boots. Each image has the same size and format as MNIST ($28 \times 28$ pixels), facilitating fair comparison across models.

CIFAR-10 is a widely adopted color image classification dataset comprising 50,000 training images and 10,000 testing images, each with dimensions of $32 \times 32$ pixels. The images are distributed evenly across 10 classes representing common objects such as airplanes, automobiles, birds, and cats. CIFAR-10 provides a more complex and diverse challenge due to its colored images and greater intra-class variability.

For model evaluation, we employ three popular neural network architectures: a Convolutional Neural Network (CNN)~\cite{chua1997cnn}, ResNet~\cite{he2016deep}, and AlexNet~\cite{krizhevsky2012imagenet}. Detailed specifications of the model structures are presented in Table~\ref{tab:model_structure}.

\begin{table}[!t]
  \centering
  \caption{Model Structures Used in Our Experiments}
  \label{tab:model_structure}
  \renewcommand{\arraystretch}{1}
  \setlength{\tabcolsep}{8pt}
  \begin{tabular}{c l p{4.6cm}}
    \toprule
    \textbf{Model} & \textbf{Dataset} & \textbf{Model Structure} \\
    \midrule
    \multirow{3}{*}{ResNet}
      & MNIST           & Conv(1→16), 3×ResBlock(16→32→64), AvgPool, FC(64→10) \\
      & Fashion-MNIST   & Conv(1→16), 3×ResBlock(16→32→64), AvgPool, FC(64→10) \\
      & CIFAR-10        & Conv(3→16), 3×ResBlock(16→32→64), AvgPool, FC(64→10) \\
    \midrule
    \multirow{3}{*}{AlexNet}
      & MNIST           & 6×Conv, 2×MaxPool, FC(4608→2048→1024→10) \\
      & Fashion-MNIST   & 6×Conv, 2×MaxPool, FC(4608→2048→1024→10) \\
      & CIFAR-10        & 5×Conv, 3×MaxPool, FC(1024→4096→4096→10) \\
    \midrule
    \multirow{3}{*}{CNN}
      & MNIST           & 2×Conv, 2×MaxPool, FC(320→50→10) \\
      & Fashion-MNIST   & 2×Conv, 2×MaxPool, FC(320→50→10) \\
      & CIFAR-10        & 2×Conv, 2×MaxPool, FC(500→50→10) \\
    \bottomrule
  \end{tabular}
\end{table}

{\bf Metrics.} We evaluate the performance of our proposed scheme using two standard metrics widely adopted in machine learning research: \textbf{model accuracy} and \textbf{inference time}. 

\textit{Model Accuracy(Acc):} This metric measures the proportion of correctly predicted labels over the total number of samples in the test set. It provides an intuitive and interpretable indicator of the effectiveness of the trained model. 

\textit{Inference Time(IT):} Inference time refers to the number of training epochs required for the global model to reach a predefined target accuracy threshold $\tau$. It reflects the convergence efficiency of the training process. A smaller inference time indicates faster convergence under the same experimental conditions.

Formally, let $A^{(t)}$ be the test accuracy at epoch $t$. The inference time $T_{\text{inf}}$ is defined as:
\[
T_{\text{inf}} = \min \left\{ t \in \mathbb{N} \,\middle|\, A^{(t)} \geq \tau \right\}.
\]

{\bf Baseline.} We conduct three groups in this experiment, including Fedavg, our proposed ACuMPA attack scheme, and the ACuMPA attack with EByFTVeS scheme.

\subsubsection{Experimental Results}
\begin{figure*}[!ht]
    \centering
    \setlength{\tabcolsep}{2pt}
    \renewcommand{\arraystretch}{0.0}

    \subfloat[MNITST+CNN]{
        \begin{tabular}{cccc}
            \includegraphics[width=0.25\textwidth,trim=70 30 70 50,clip]{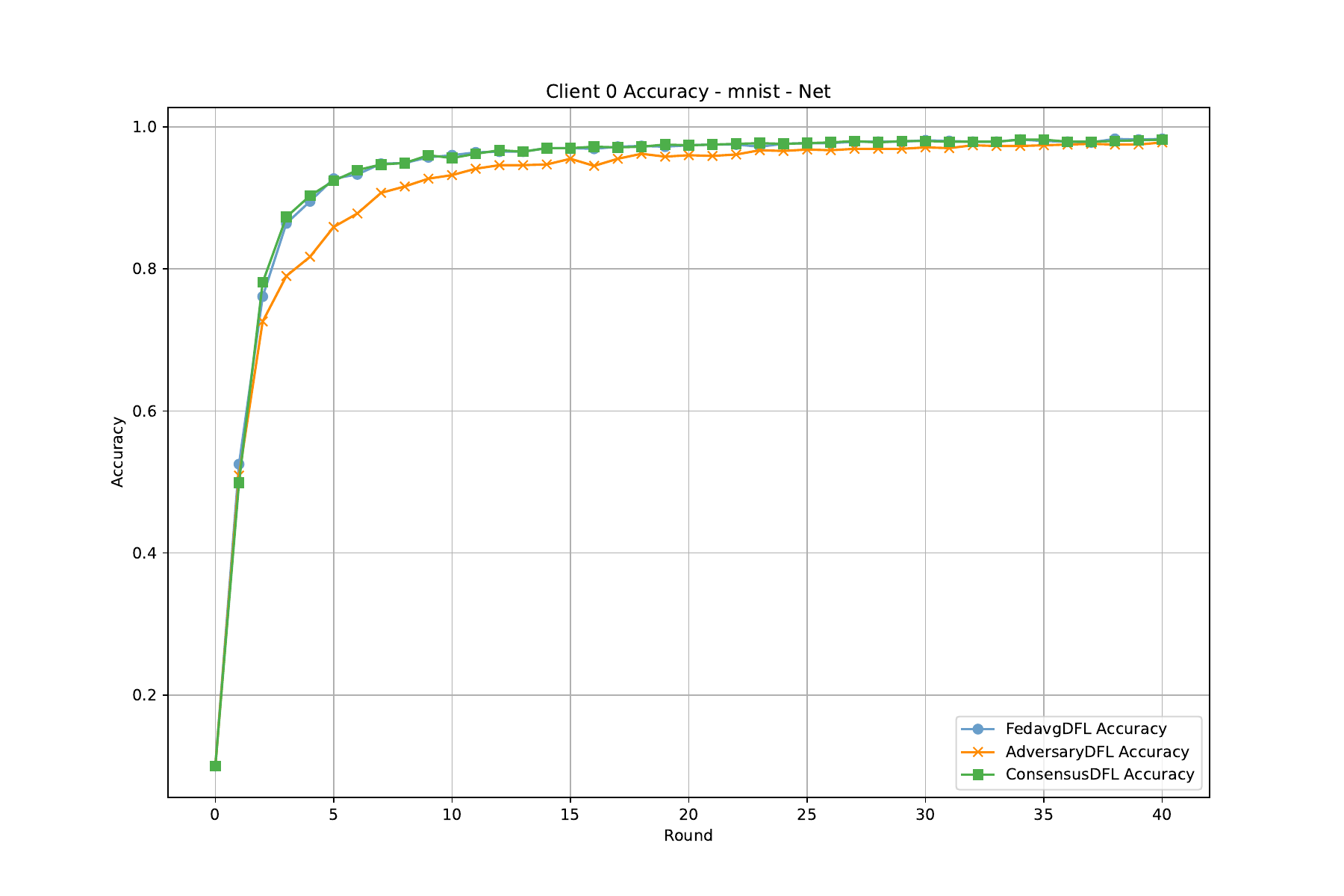} &
            \includegraphics[width=0.25\textwidth,trim=70 30 70 50,clip]{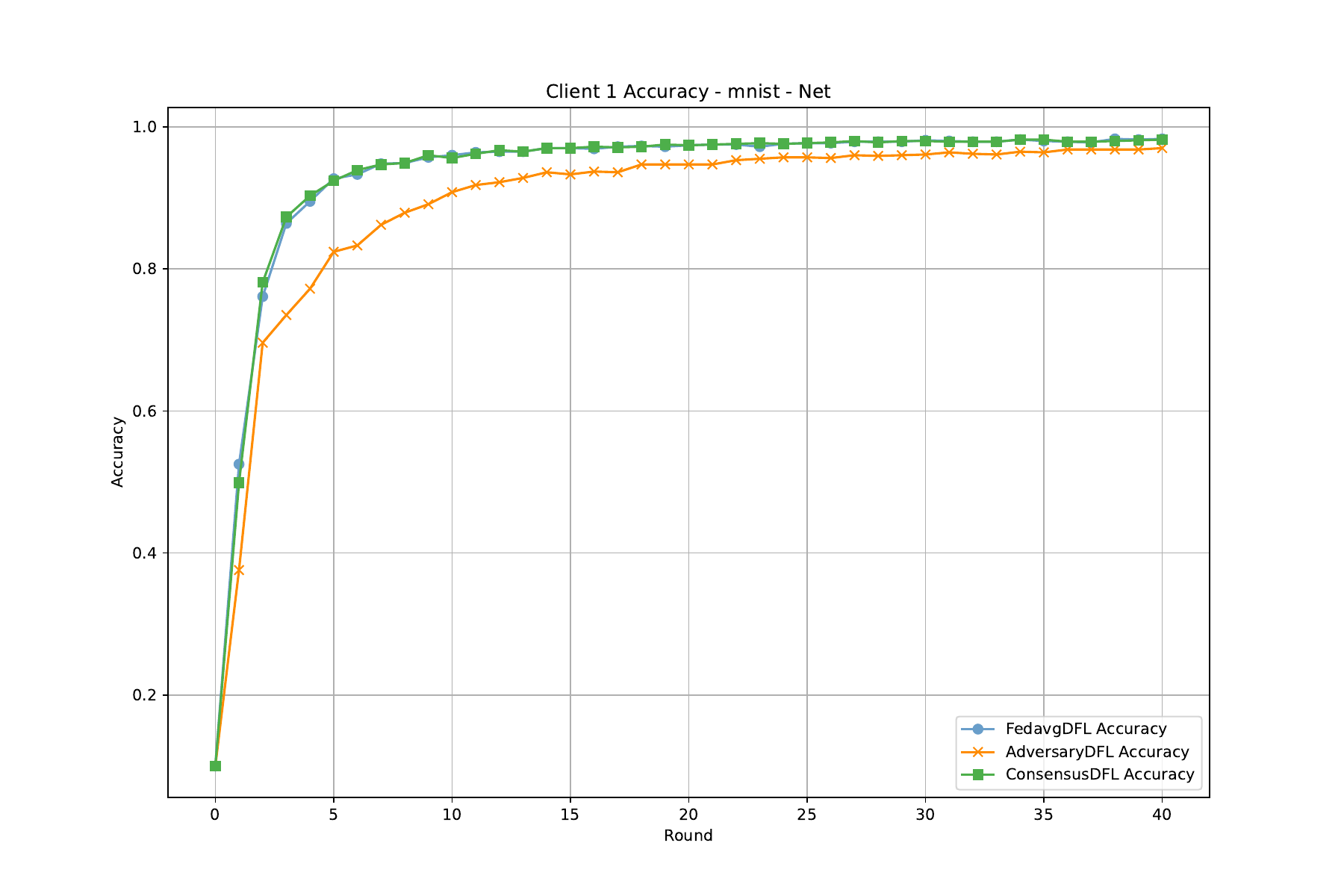} &
            \includegraphics[width=0.25\textwidth,trim=70 30 70 50,clip]{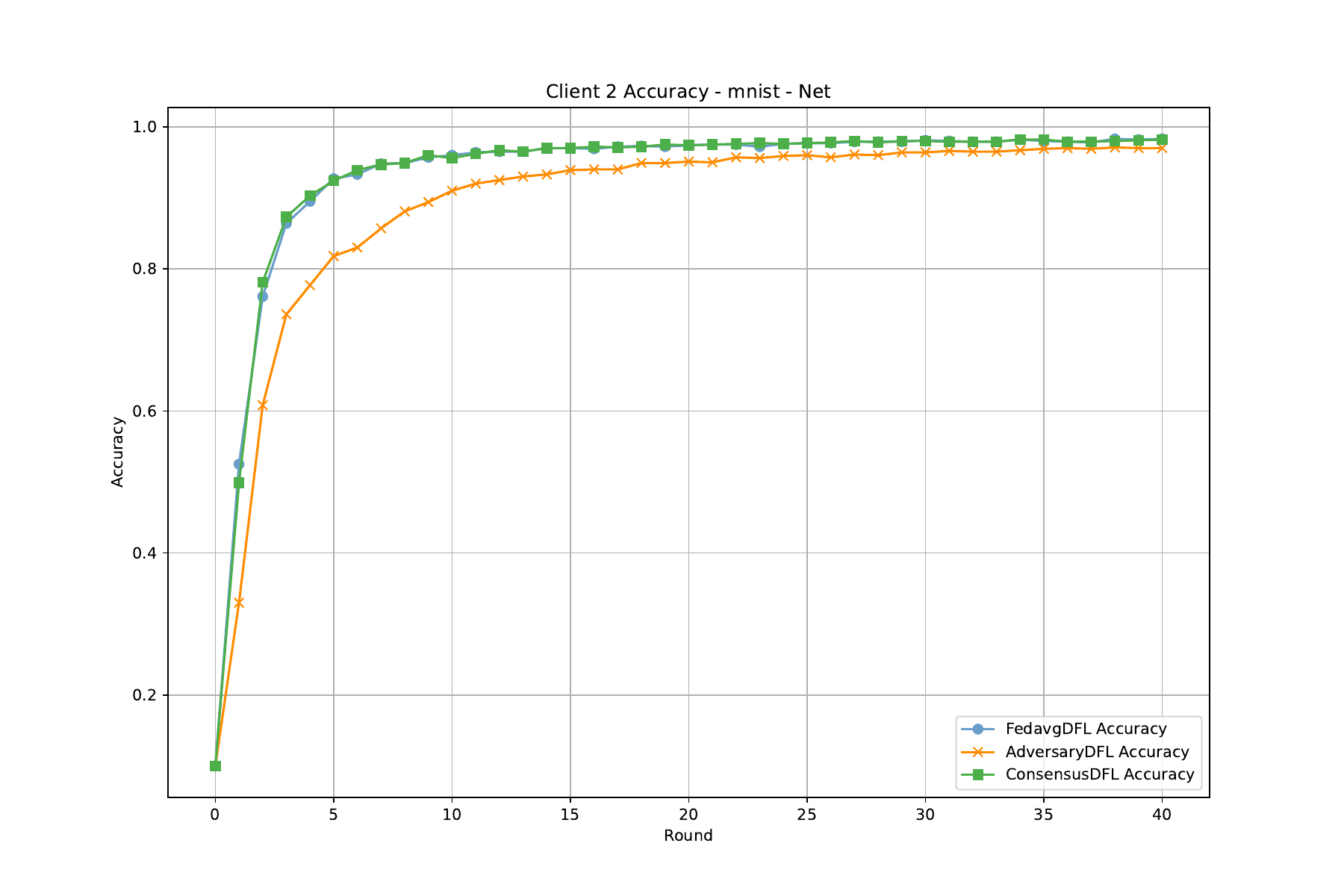} &
            \includegraphics[width=0.25\textwidth,trim=70 30 70 50,clip]{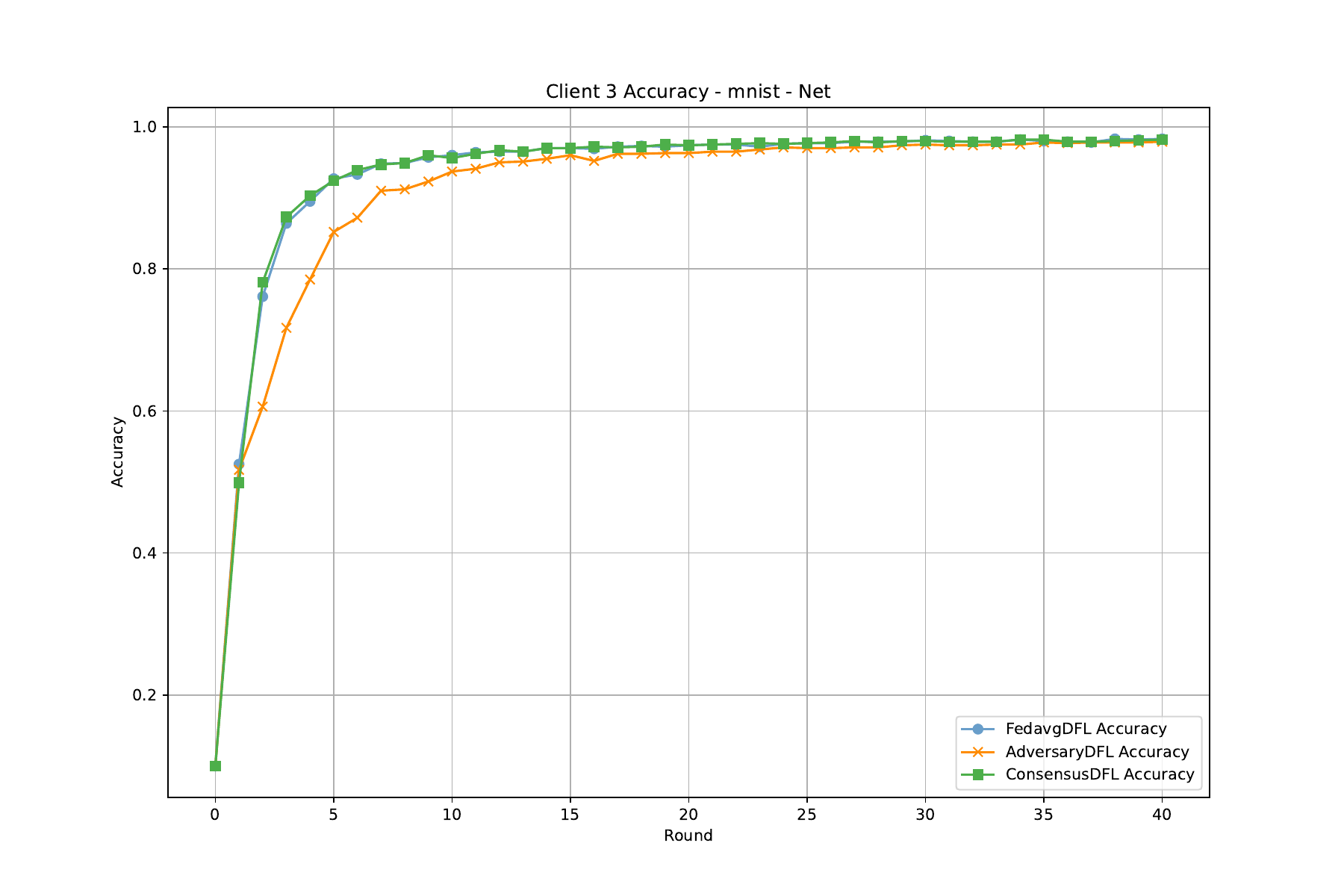}
        \end{tabular}
    }
    \vspace{0em} 
    \subfloat[Fashion-MNIST+CNN]{
        \begin{tabular}{cccc}
            \includegraphics[width=0.25\textwidth,trim=70 30 70 50,clip]{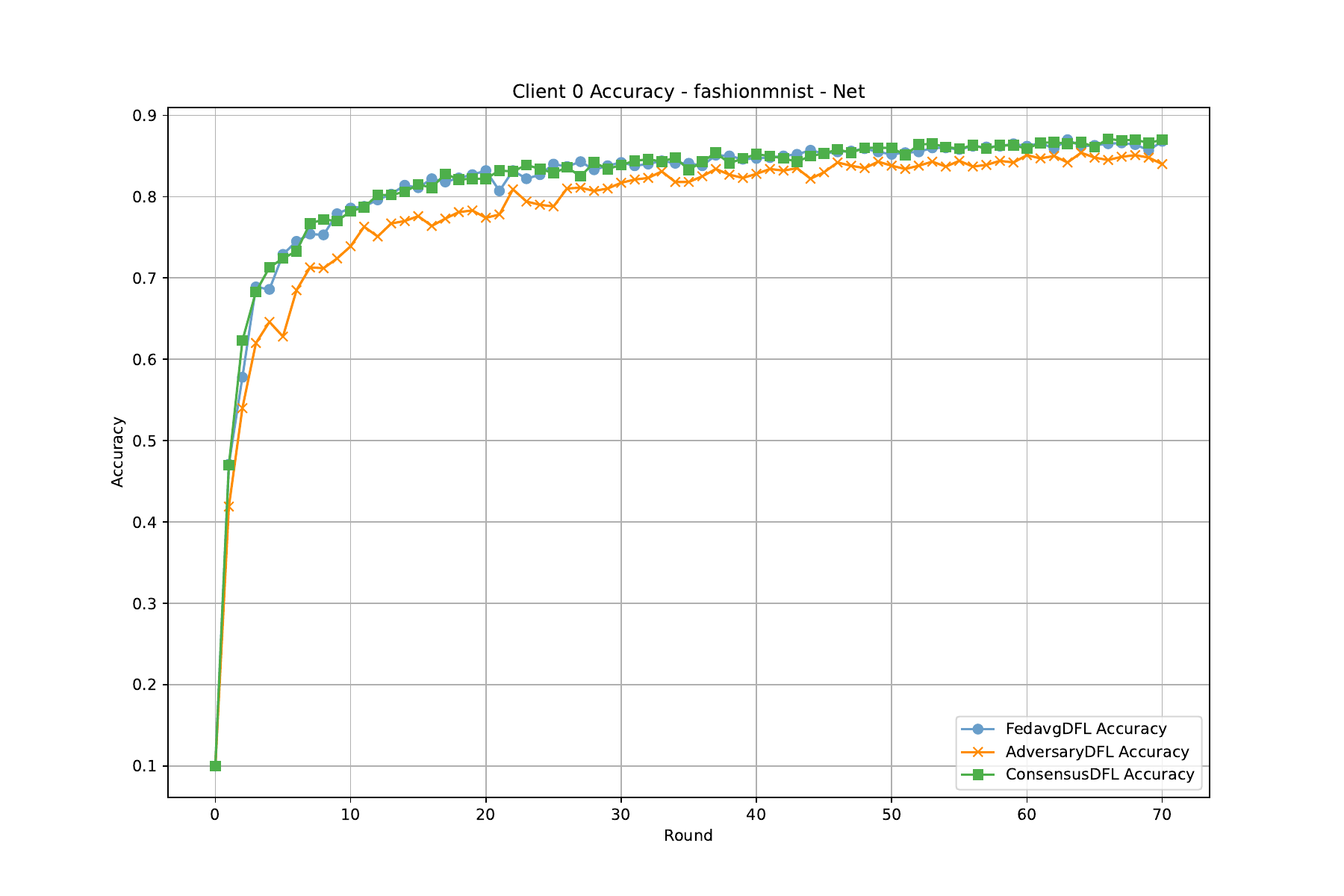} &
            \includegraphics[width=0.25\textwidth,trim=70 30 70 50,clip]{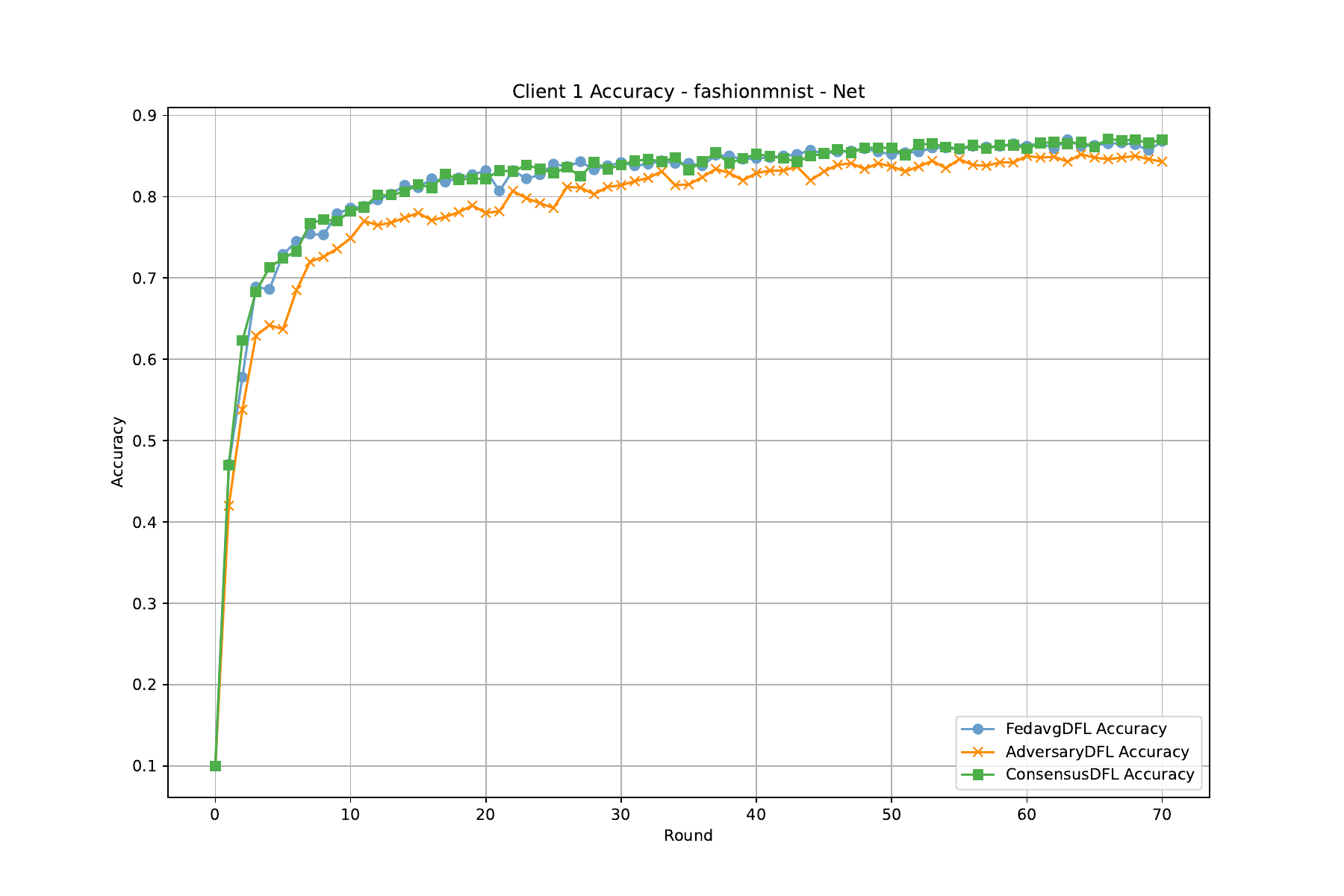} &
            \includegraphics[width=0.25\textwidth,trim=70 30 70 50,clip]{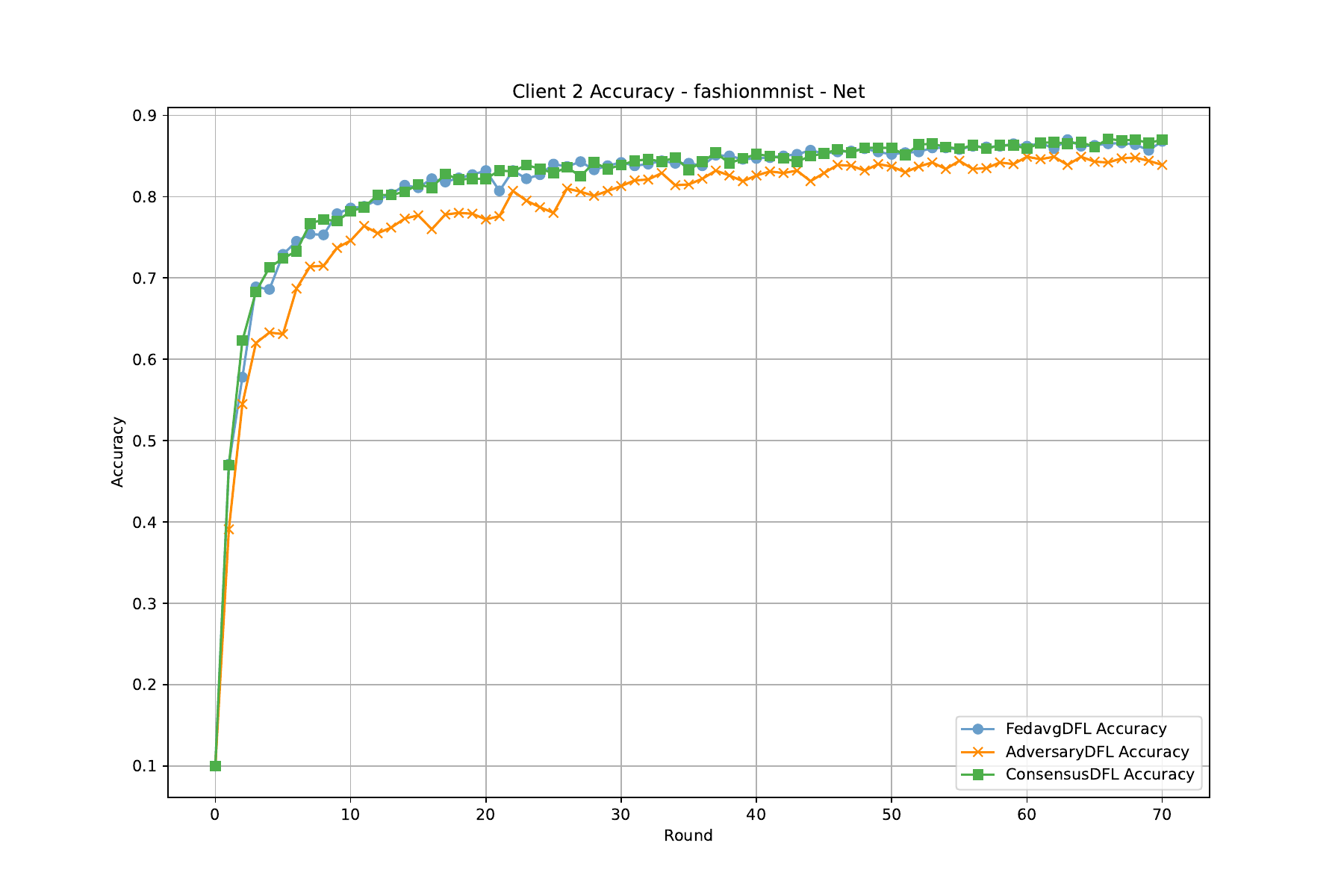} &
            \includegraphics[width=0.25\textwidth,trim=70 30 70 50,clip]{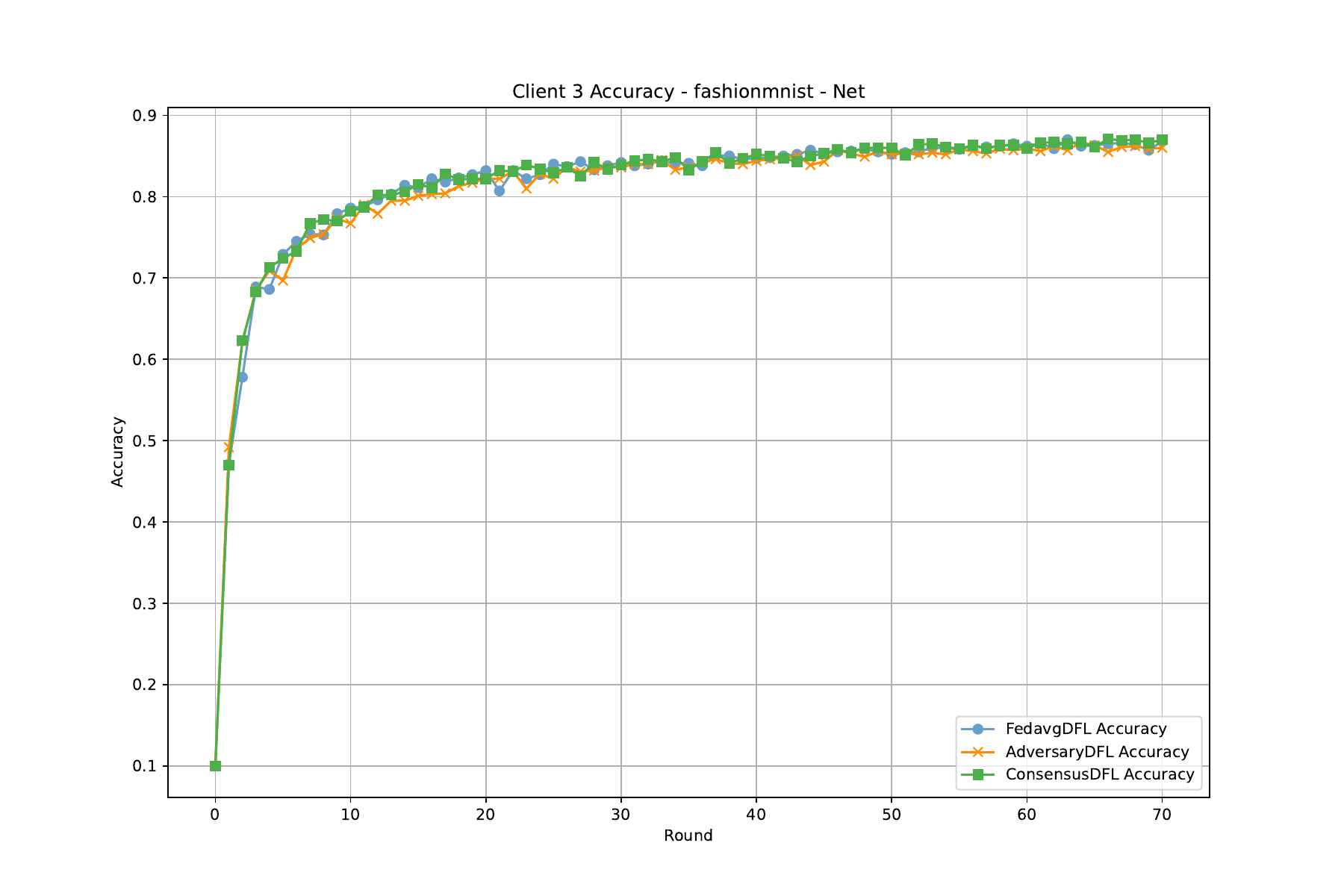}
        \end{tabular}
    }
    \vspace{0em} 
    \subfloat[CIFAR10+CNN]{
        \begin{tabular}{cccc}
            \includegraphics[width=0.25\textwidth,trim=70 30 70 50,clip]{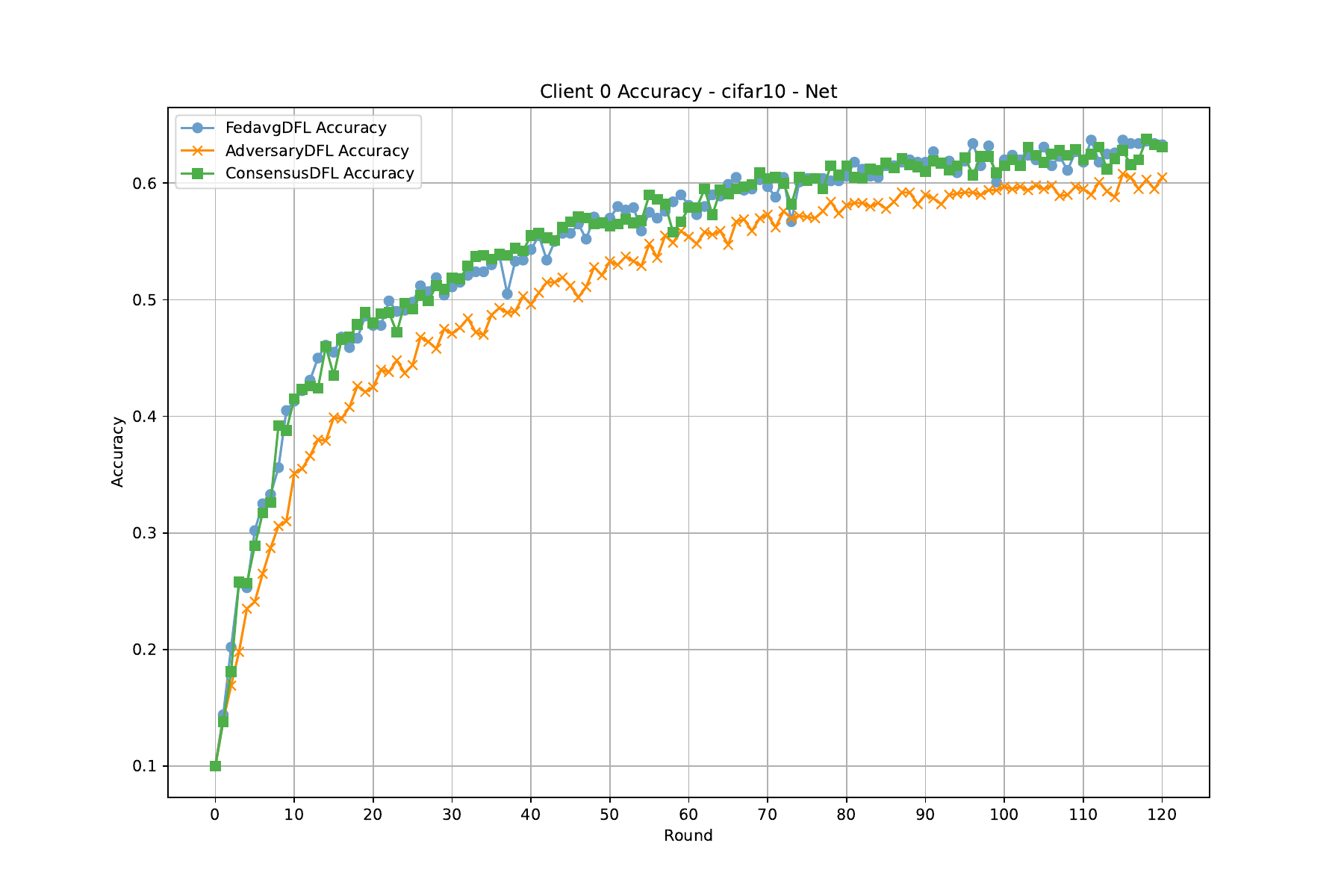} &
            \includegraphics[width=0.25\textwidth,trim=70 30 70 50,clip]{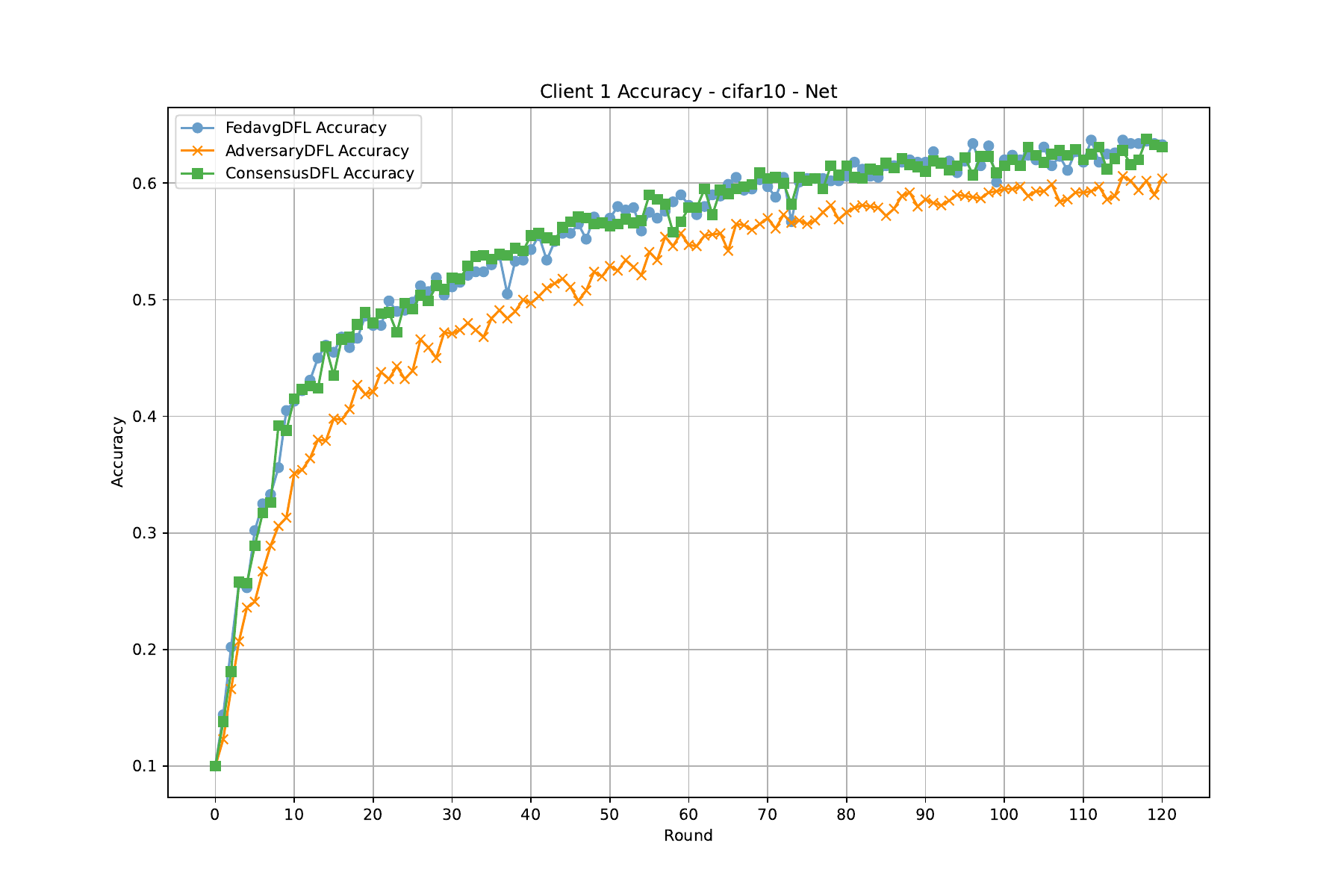} &
            \includegraphics[width=0.25\textwidth,trim=70 30 70 50,clip]{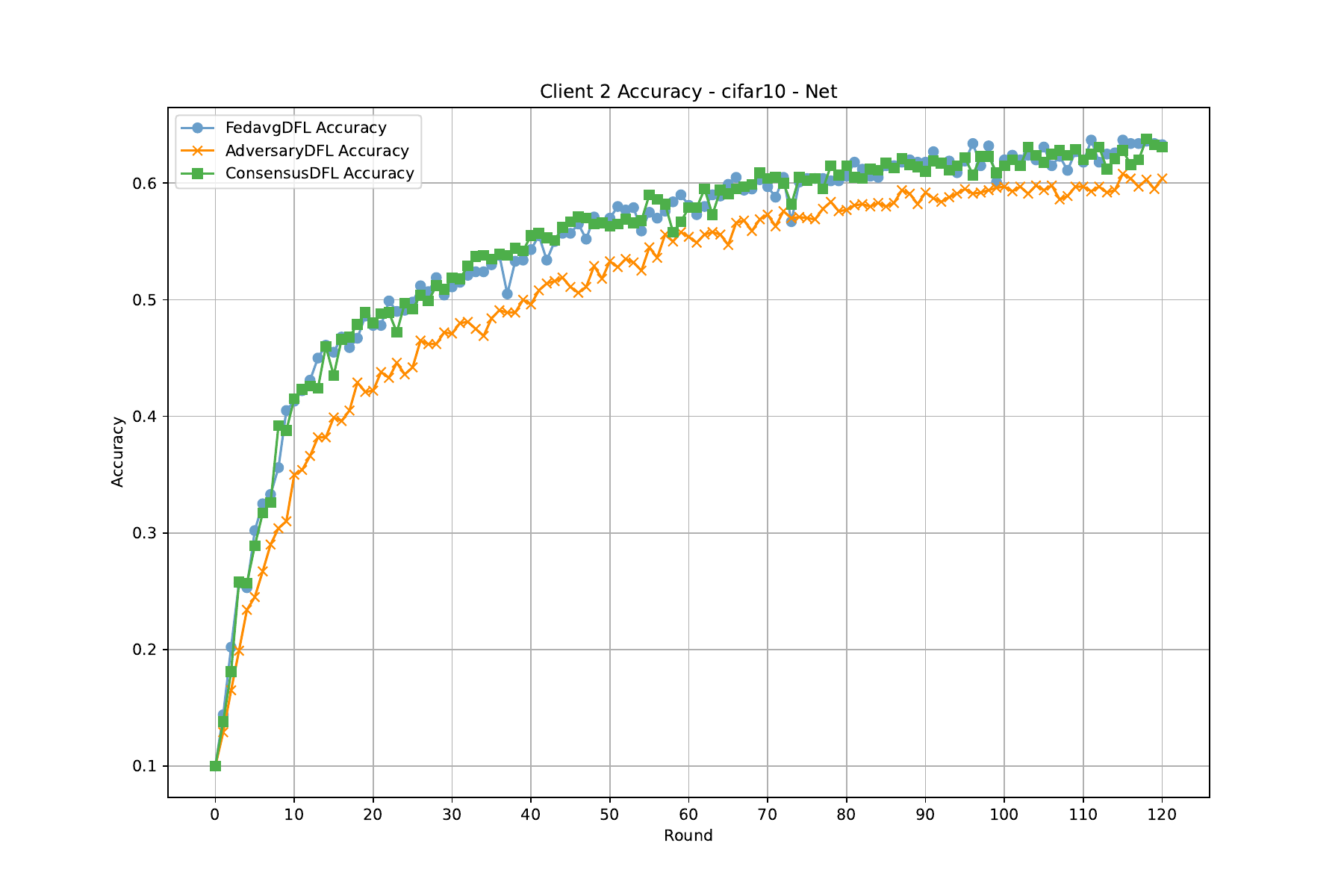} &
            \includegraphics[width=0.25\textwidth,trim=70 30 70 50,clip]{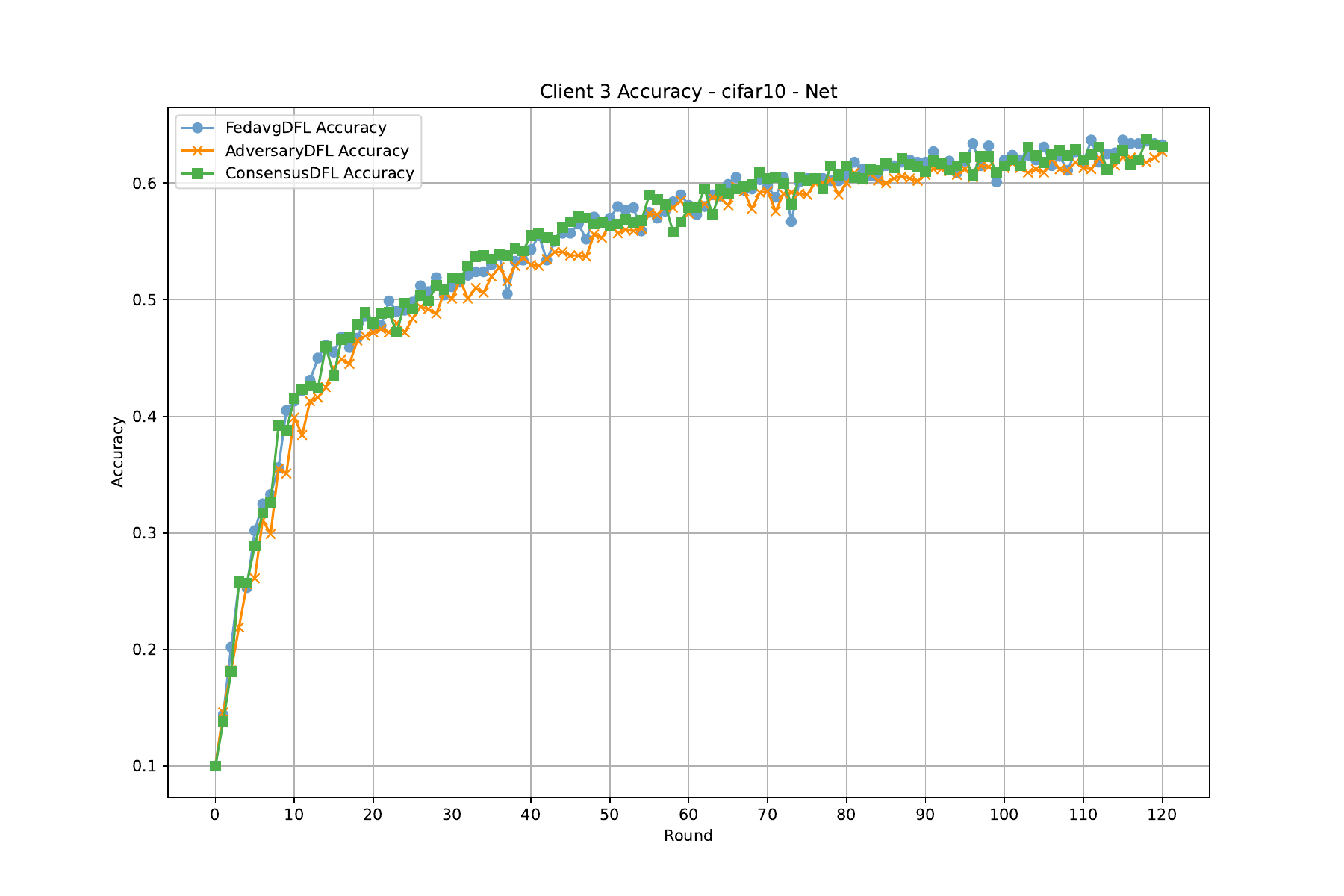}
        \end{tabular}
    }

    \caption{Accuracy Results using CNN model.}
    \label{fig:cnn_results}
\end{figure*}

\begin{figure*}[!ht]
    \centering
    \setlength{\tabcolsep}{2pt}
    \renewcommand{\arraystretch}{0.0}

    \subfloat[MNITST+ResNet]{
        \begin{tabular}{cccc}
            \includegraphics[width=0.25\textwidth,trim=70 30 70 50,clip]{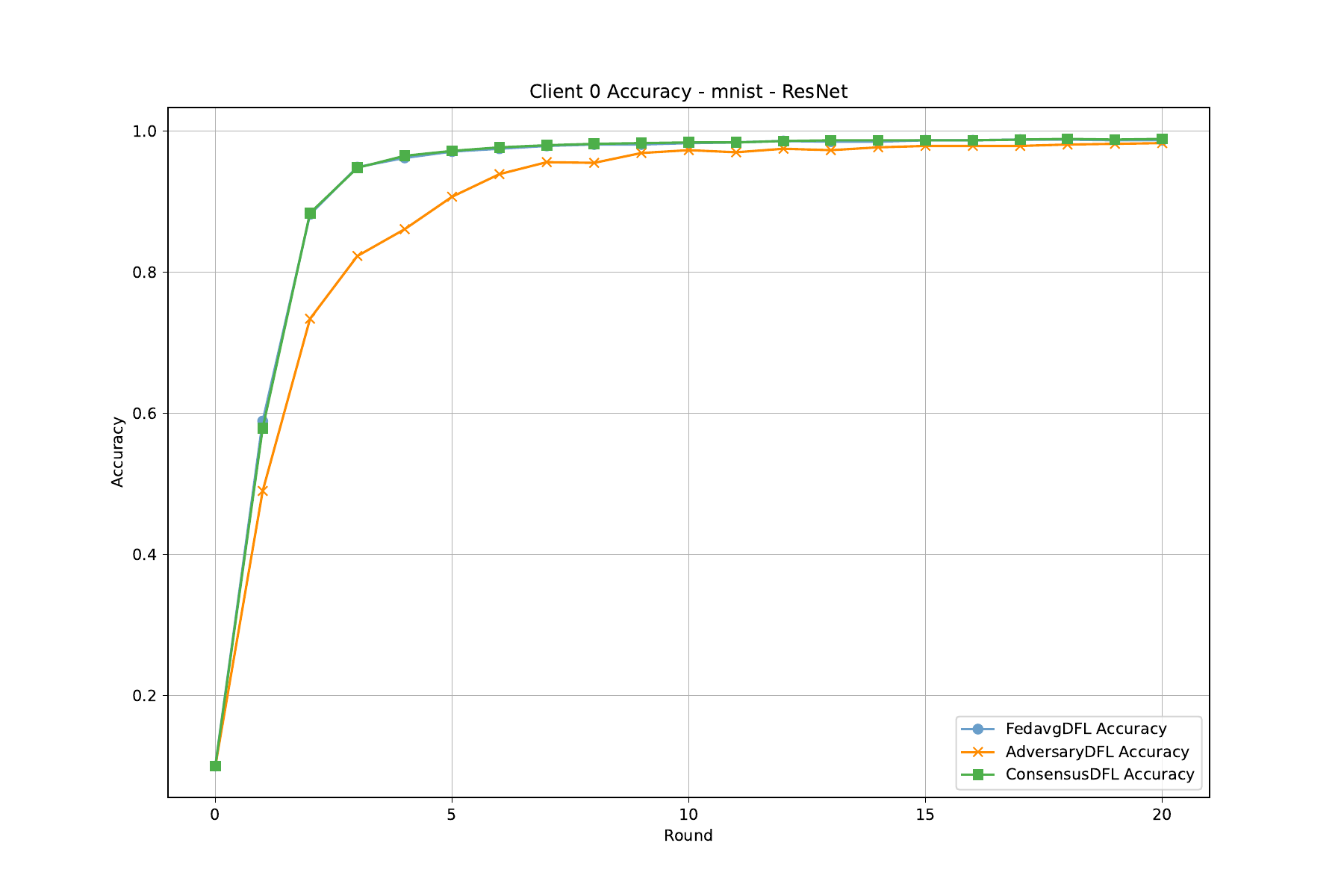} &
            \includegraphics[width=0.25\textwidth,trim=70 30 70 50,clip]{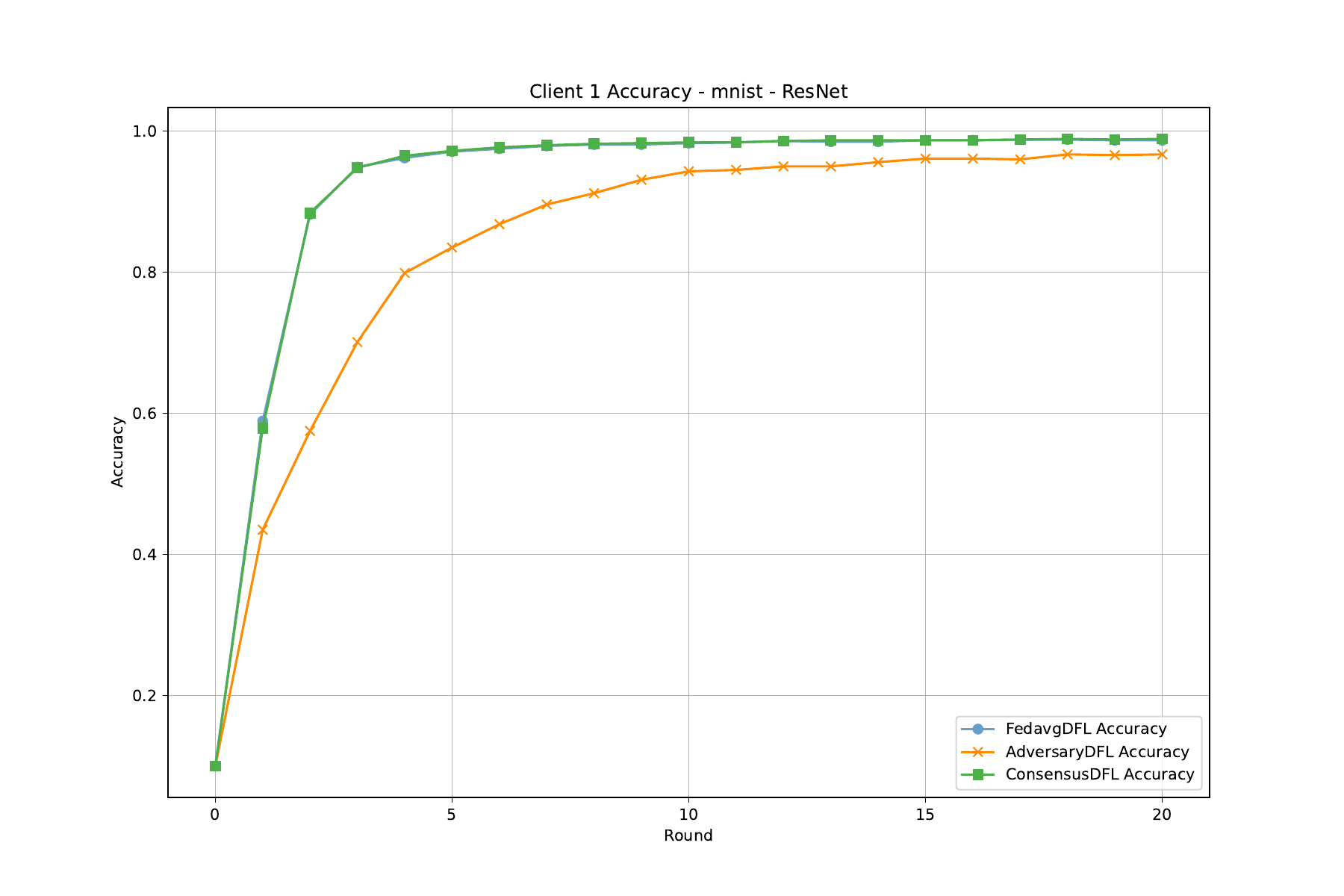} &
            \includegraphics[width=0.25\textwidth,trim=70 30 70 50,clip]{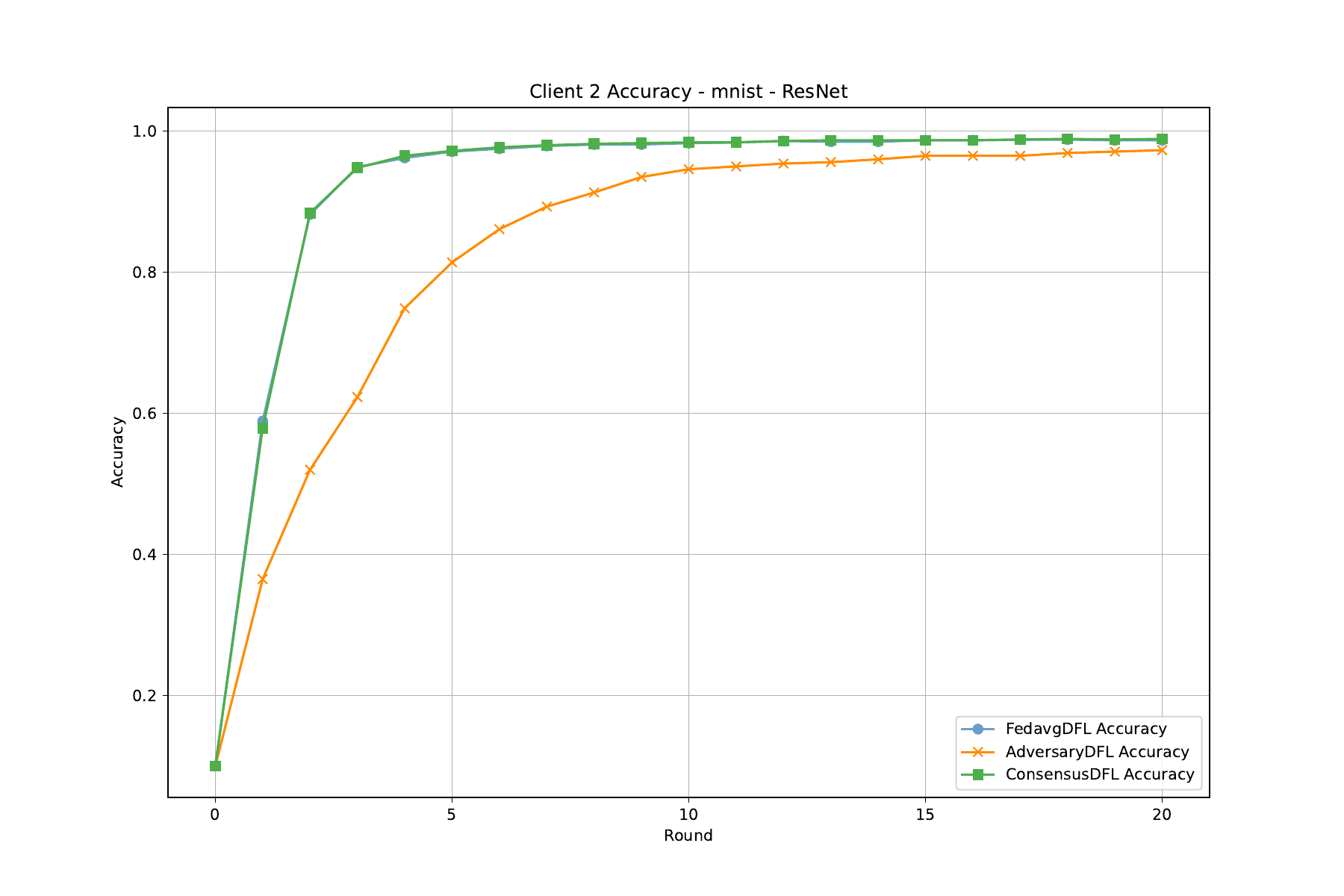} &
            \includegraphics[width=0.25\textwidth,trim=70 30 70 50,clip]{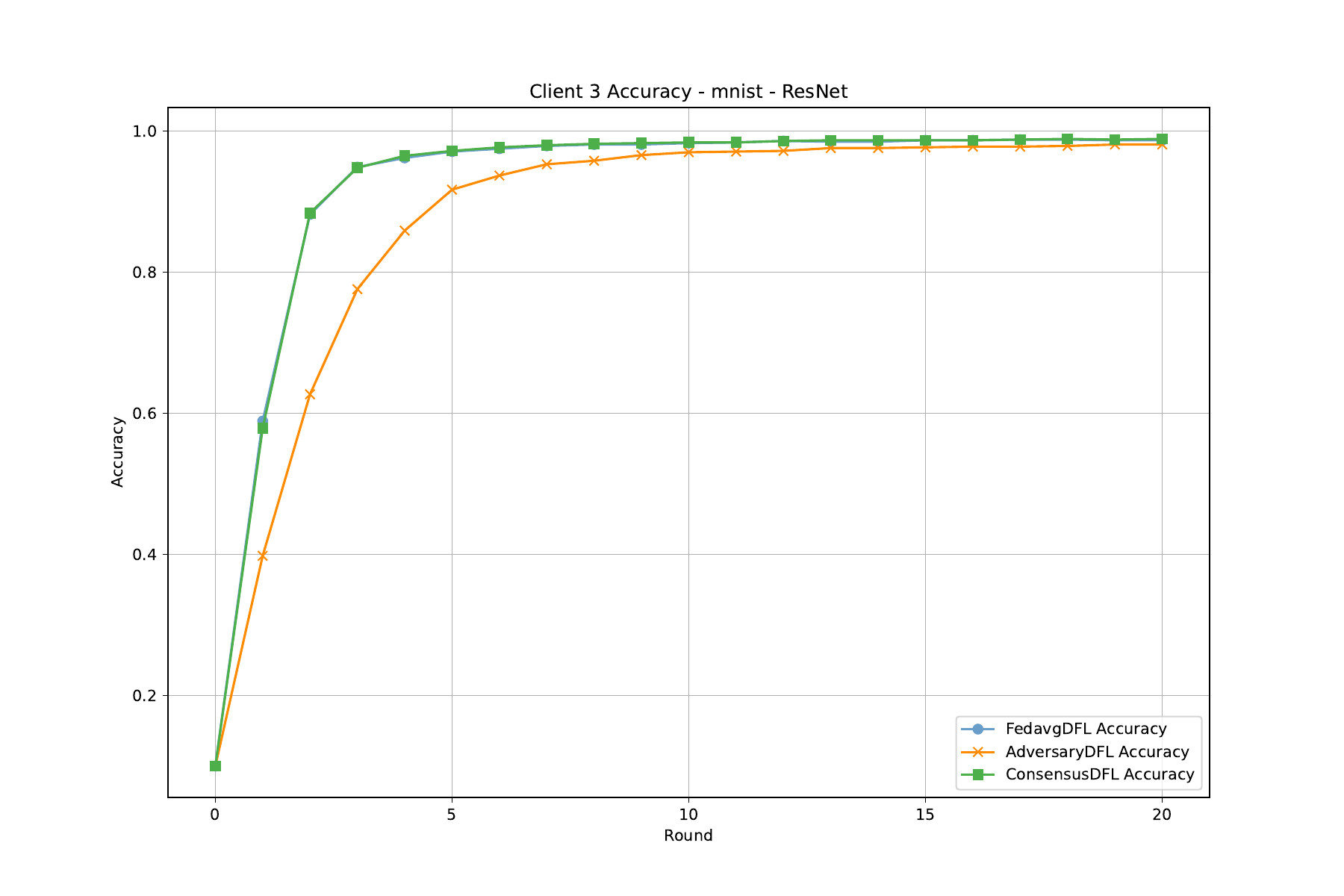}
        \end{tabular}
    }
    \vspace{0em} 
    \subfloat[Fashion-MNIST+ResNet]{
        \begin{tabular}{cccc}
            \includegraphics[width=0.25\textwidth,trim=70 30 70 50,clip]{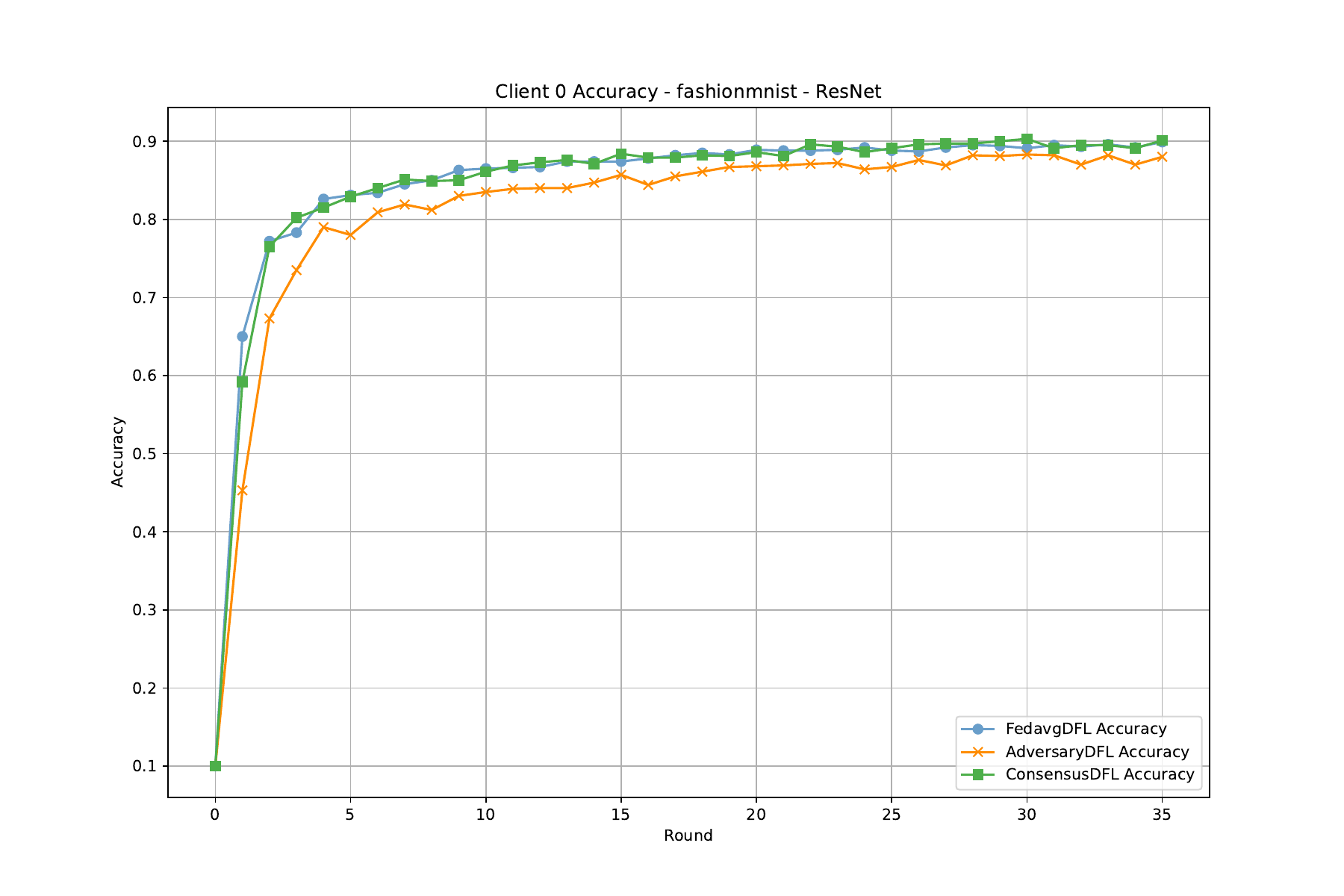} &
            \includegraphics[width=0.25\textwidth,trim=70 30 70 50,clip]{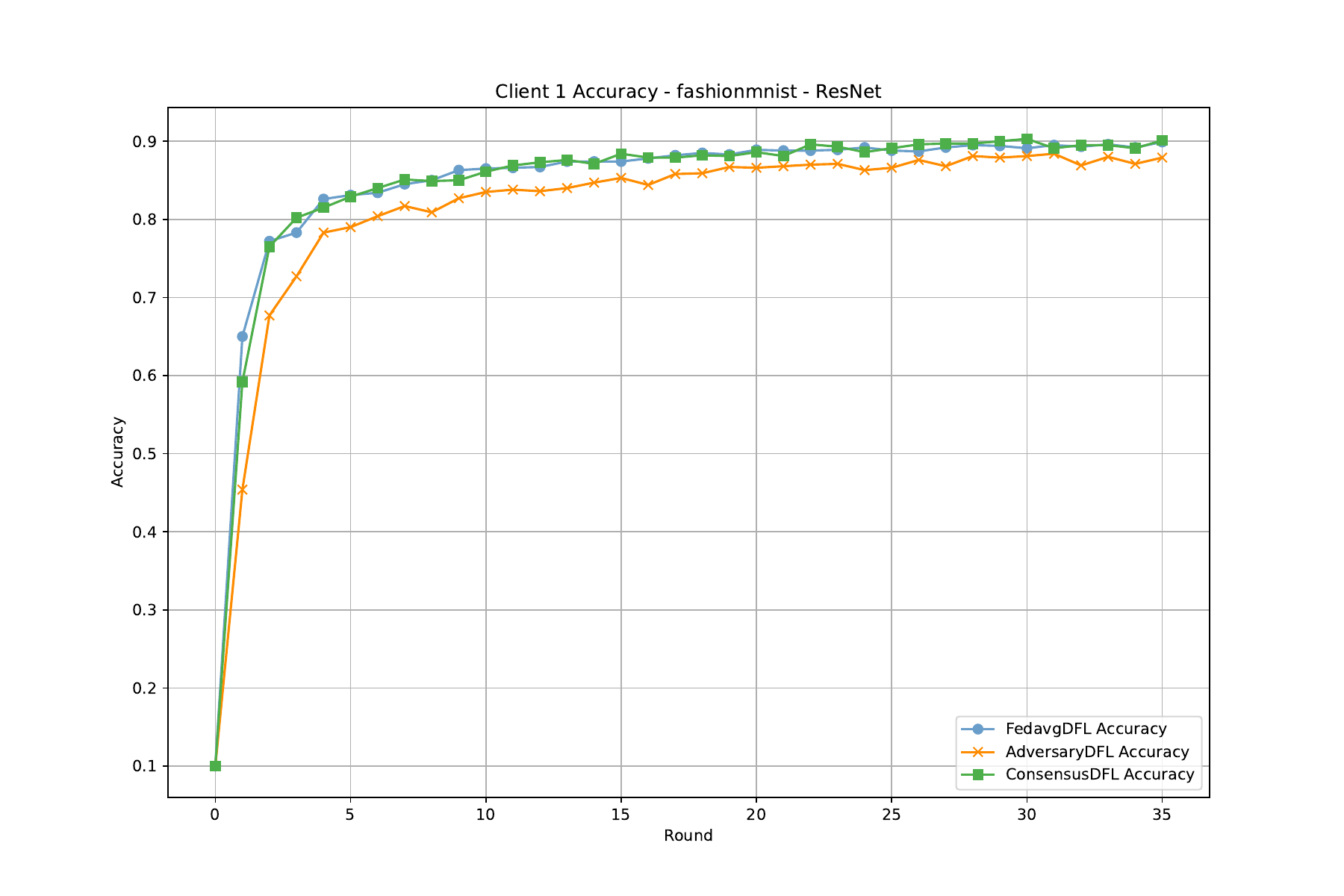} &
            \includegraphics[width=0.25\textwidth,trim=70 30 70 50,clip]{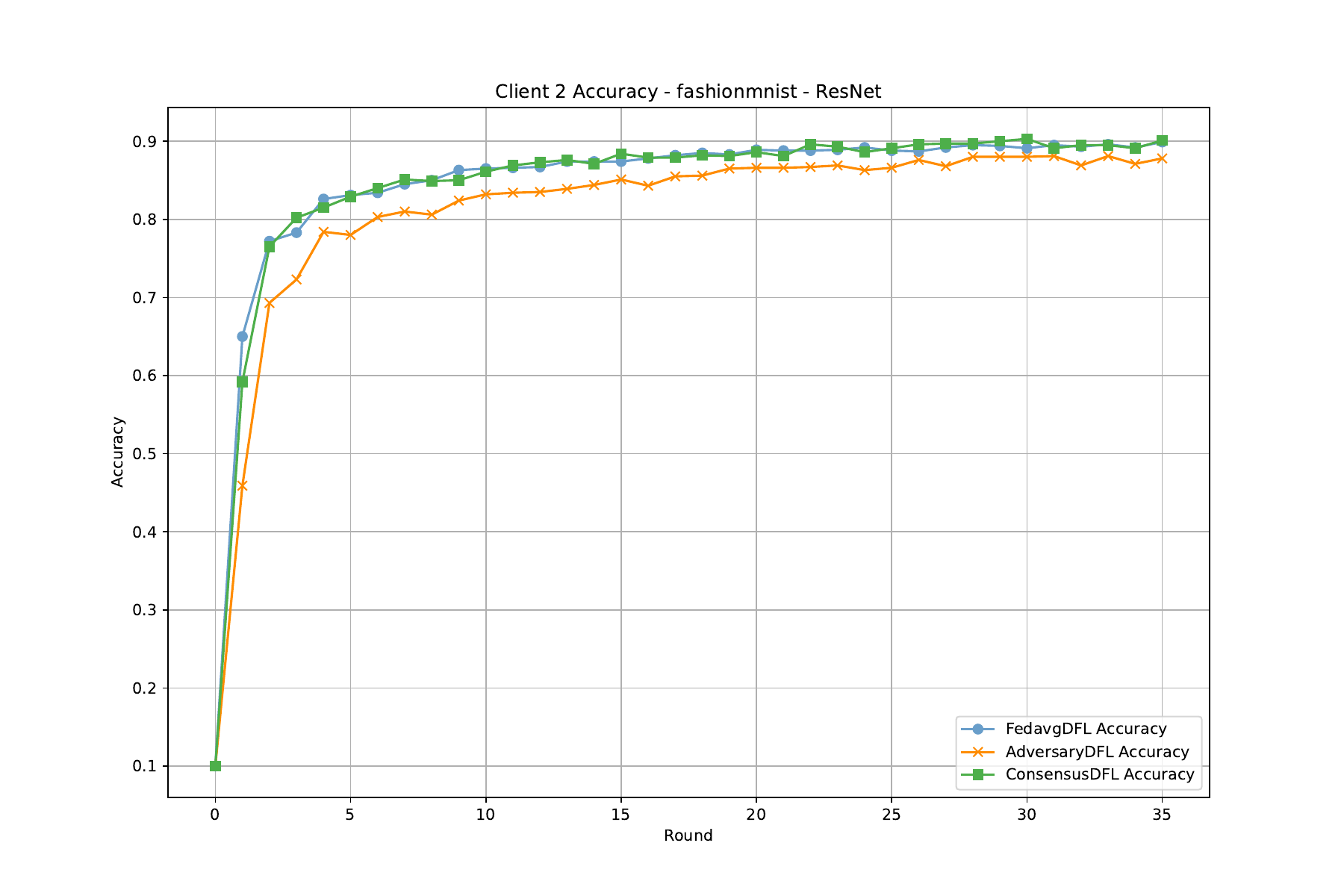} &
            \includegraphics[width=0.25\textwidth,trim=70 30 70 50,clip]{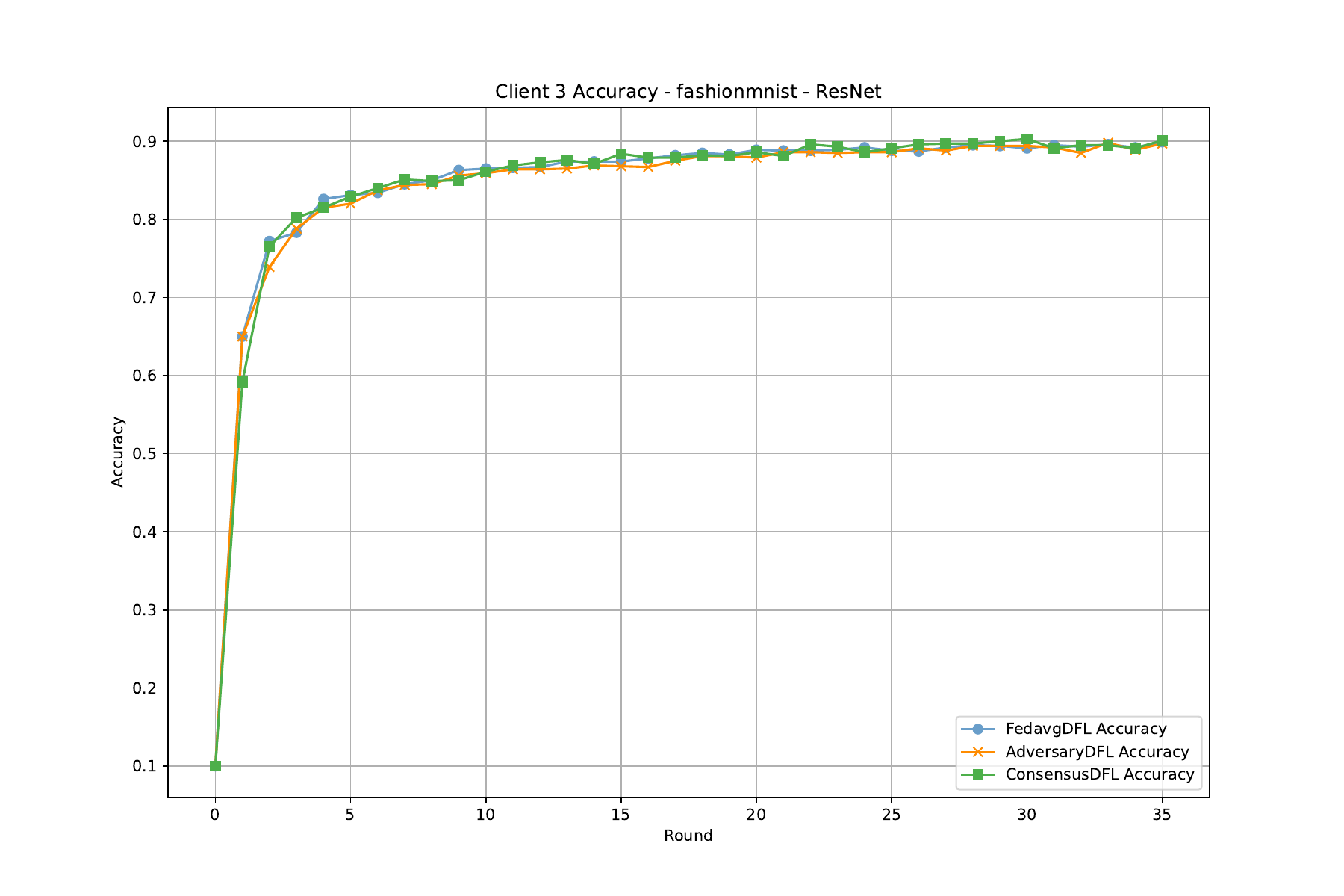}
        \end{tabular}
    }
    \vspace{0em} 
    \subfloat[CIFAR10+ResNet]{
        \begin{tabular}{cccc}
            \includegraphics[width=0.25\textwidth,trim=70 30 70 50,clip]{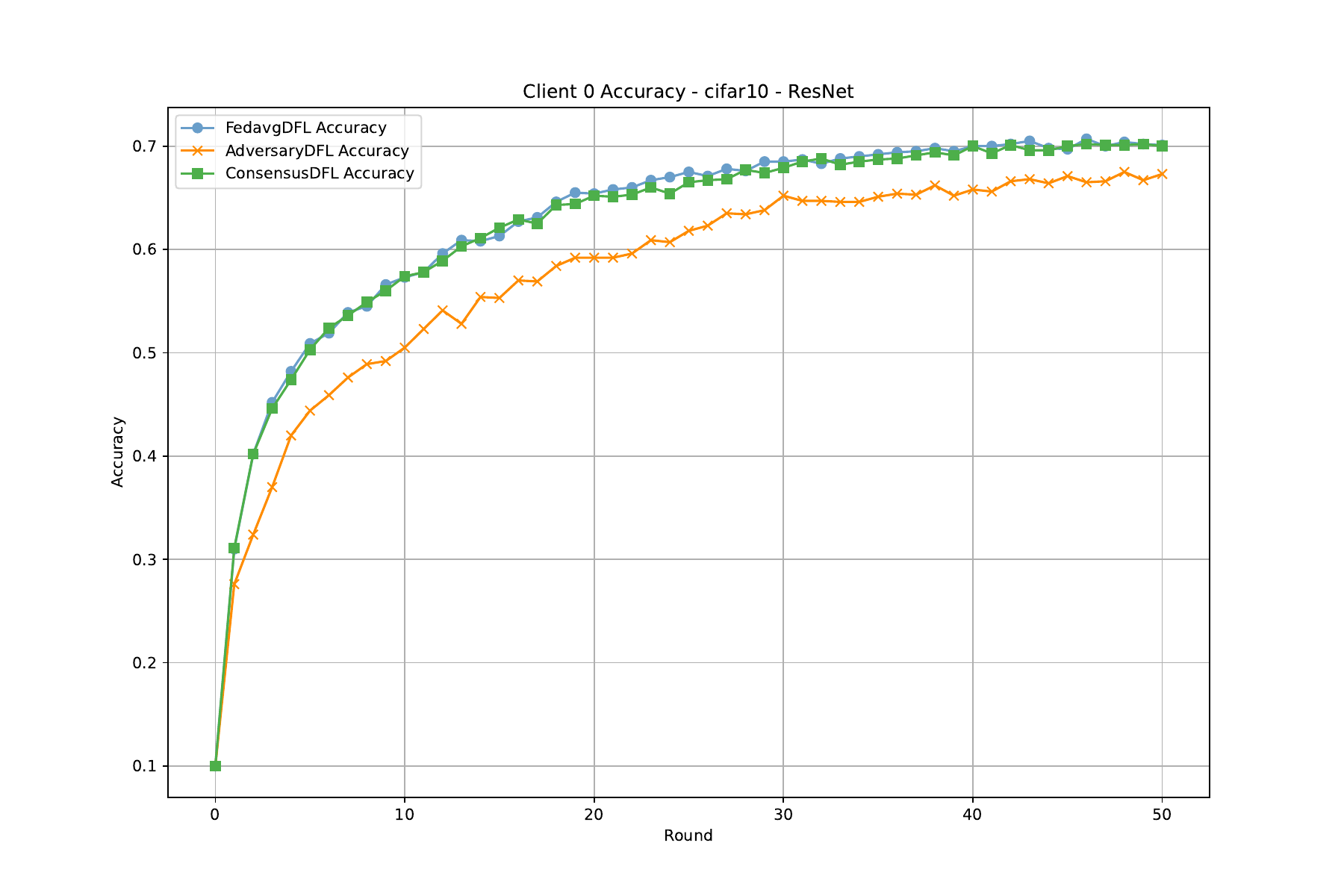} &
            \includegraphics[width=0.25\textwidth,trim=70 30 70 50,clip]{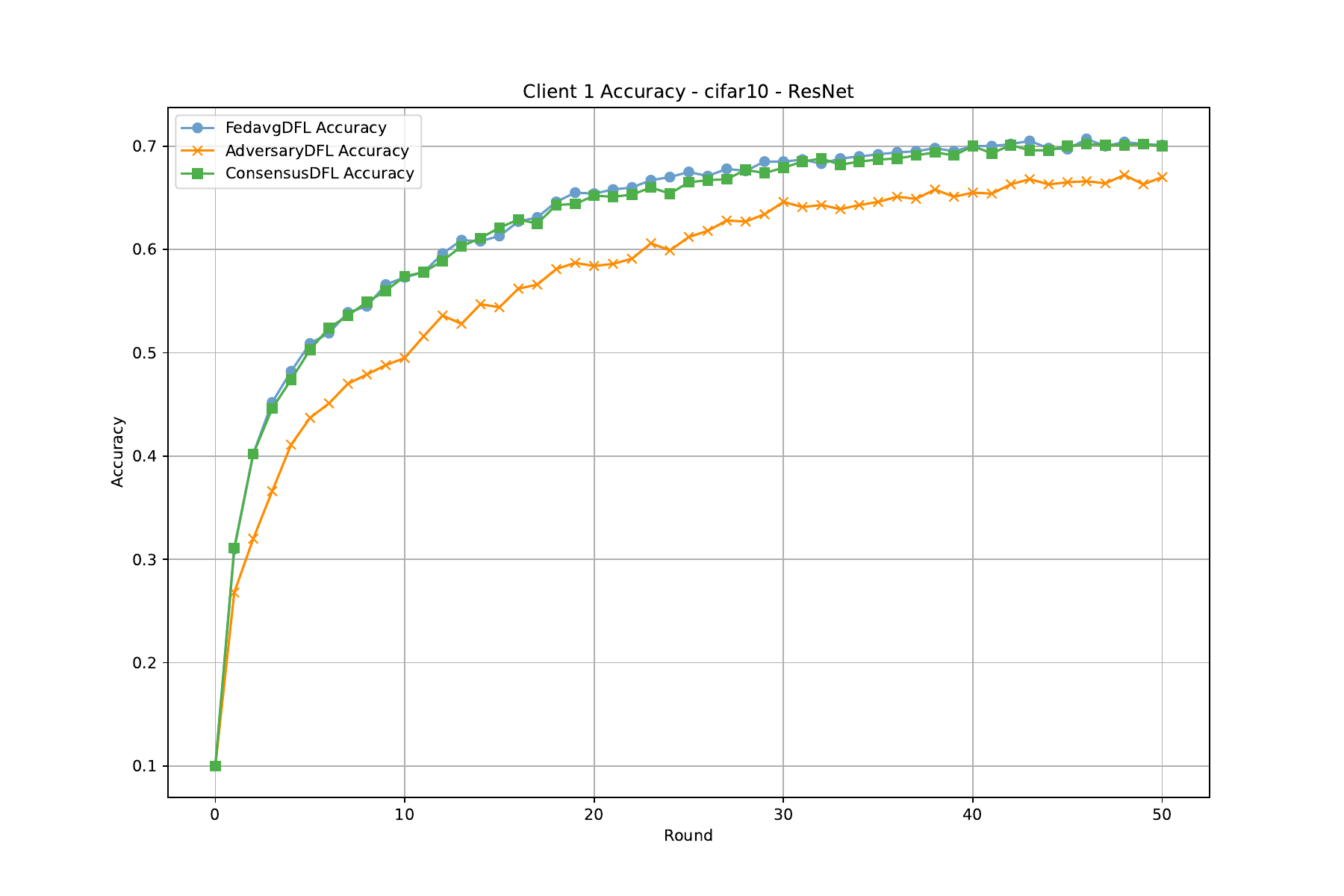} &
            \includegraphics[width=0.25\textwidth,trim=70 30 70 50,clip]{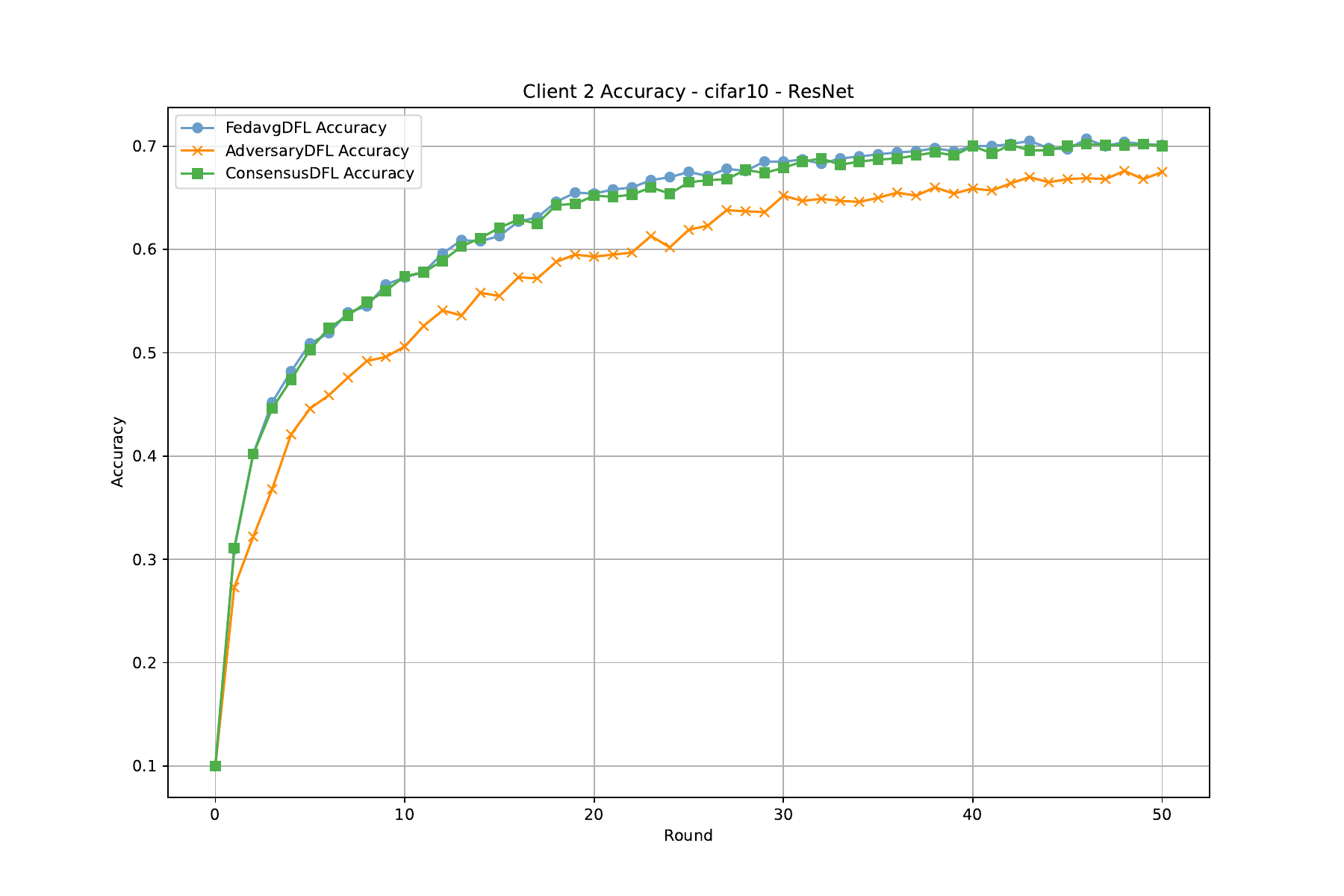} &
            \includegraphics[width=0.25\textwidth,trim=70 30 70 50,clip]{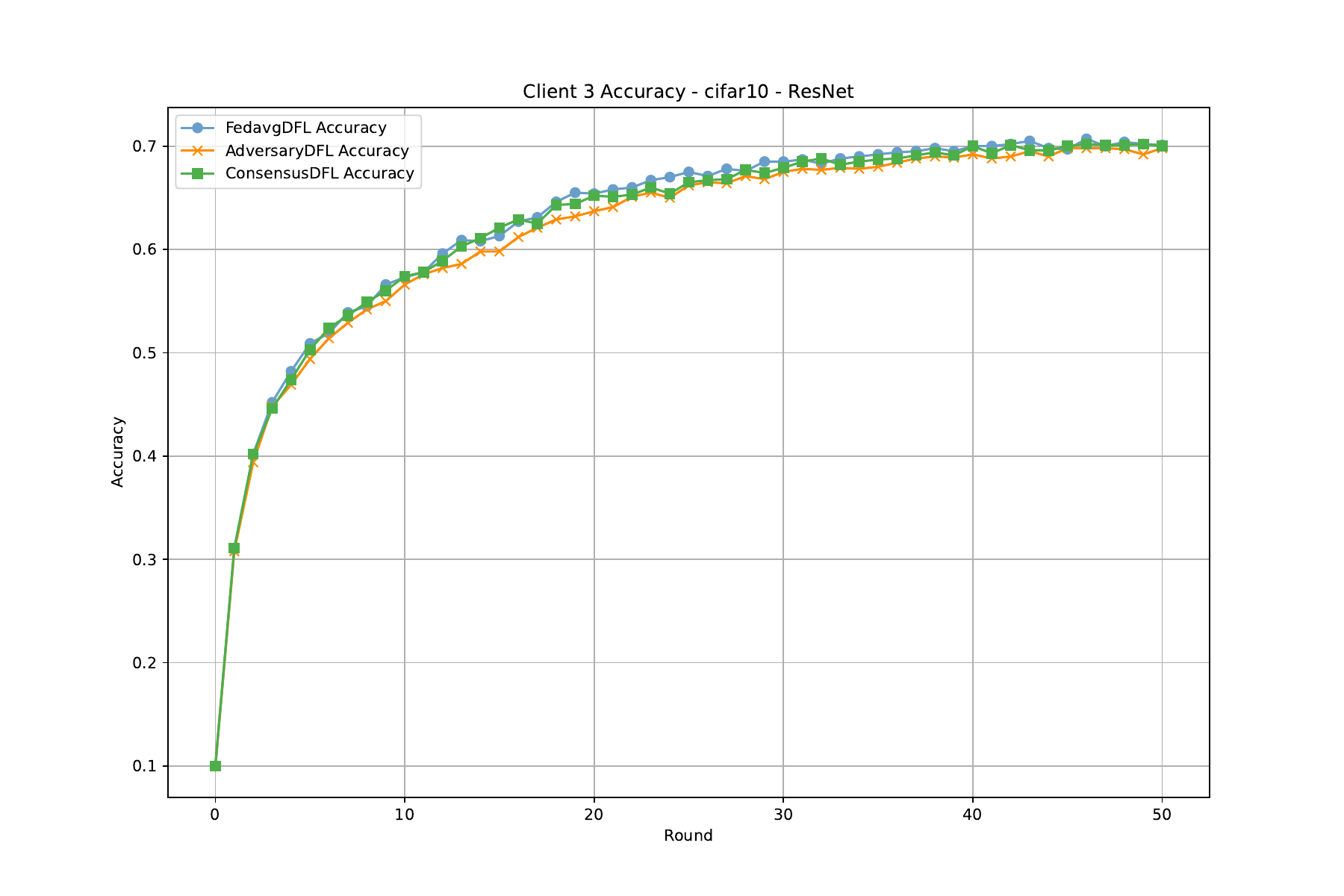}
        \end{tabular}
    }

    \caption{Accuracy Results using ResNet model.}
    \label{fig:resnet_results}
\end{figure*}

\begin{figure*}[!ht]
    \centering
    \setlength{\tabcolsep}{2pt}
    \renewcommand{\arraystretch}{0.0}

    \subfloat[MNITST+AlexNet]{
        \begin{tabular}{cccc}
            \includegraphics[width=0.25\textwidth,trim=70 30 70 50,clip]{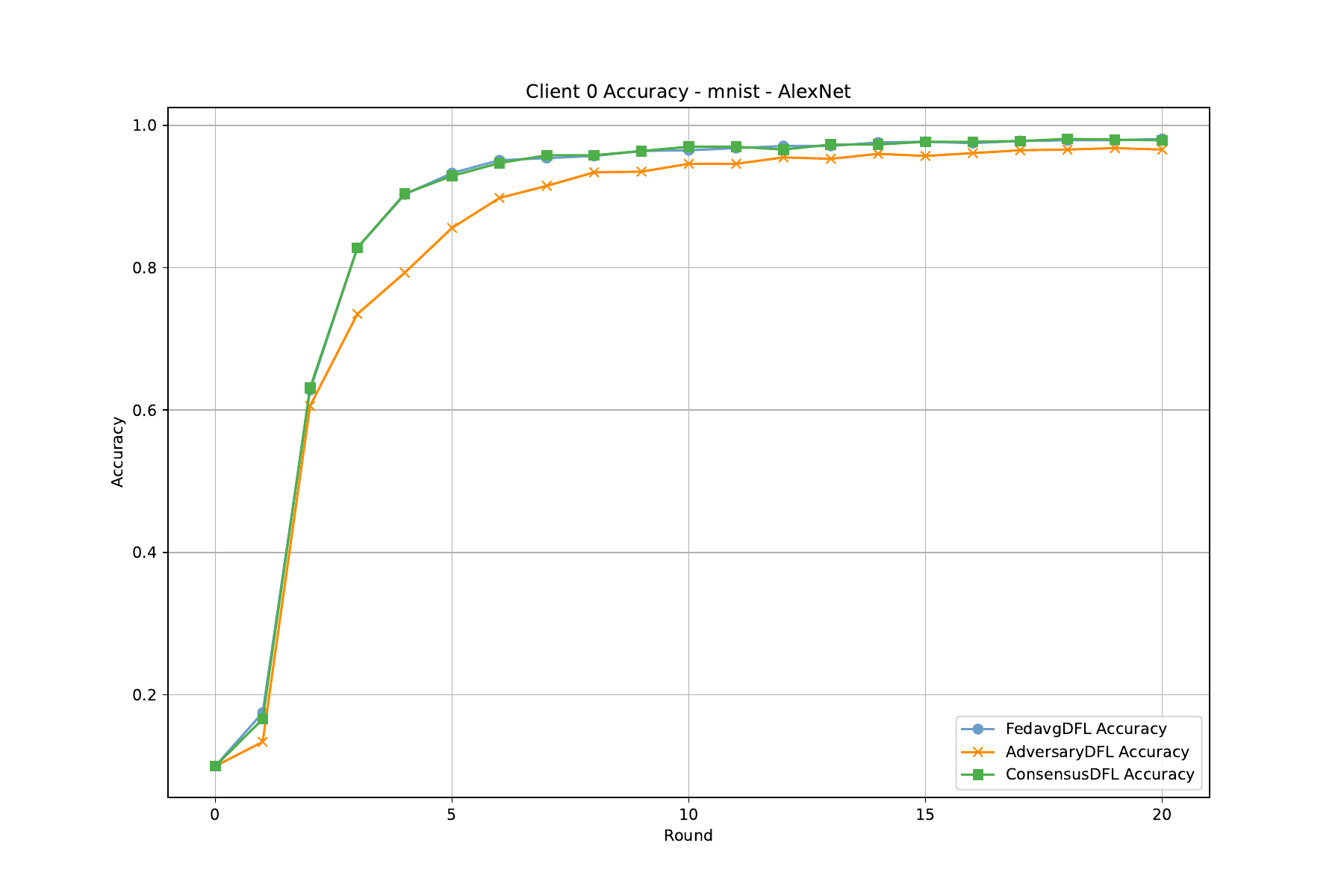} &
            \includegraphics[width=0.25\textwidth,trim=70 30 70 50,clip]{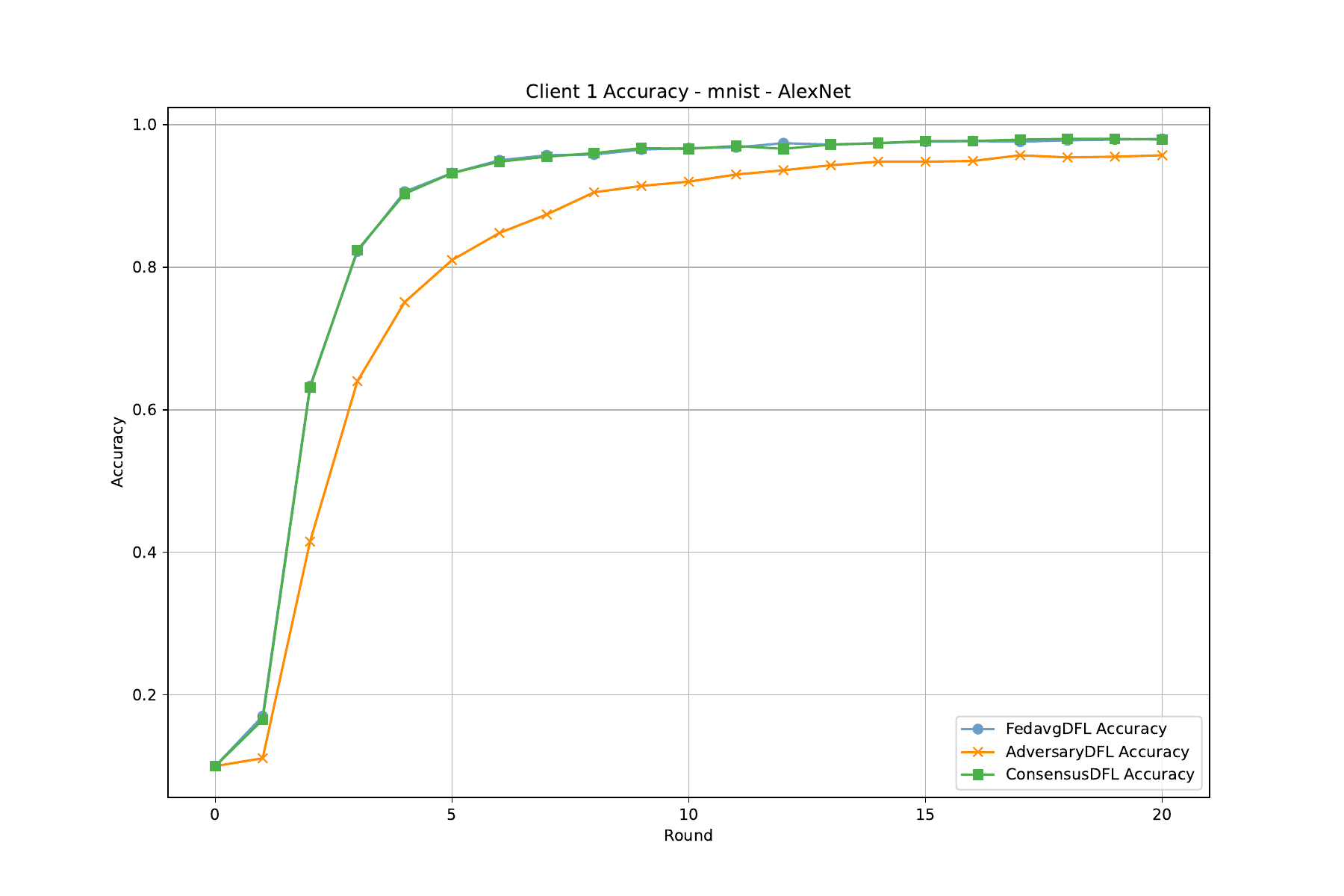} &
            \includegraphics[width=0.25\textwidth,trim=70 30 70 50,clip]{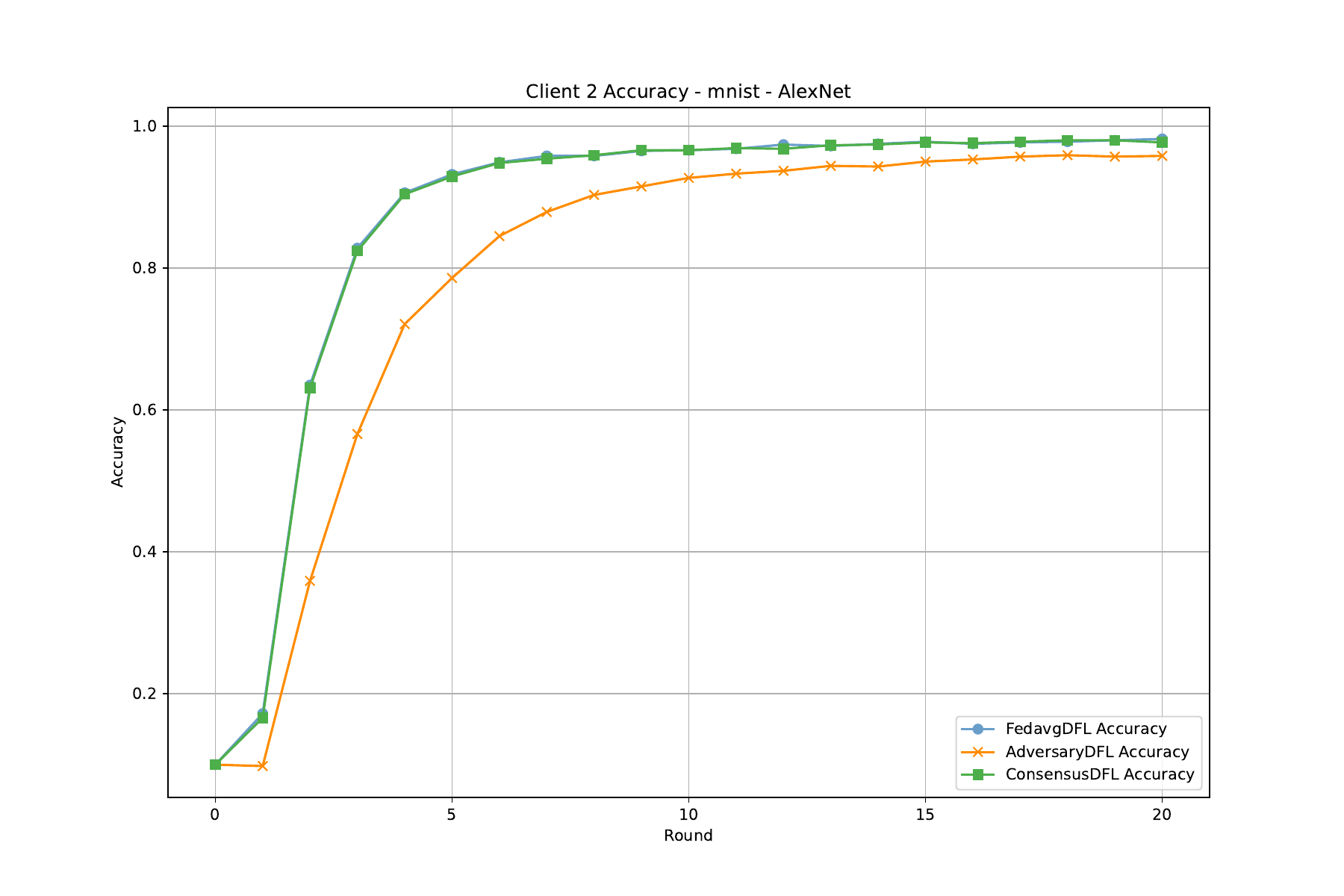} &
            \includegraphics[width=0.25\textwidth,trim=70 30 70 50,clip]{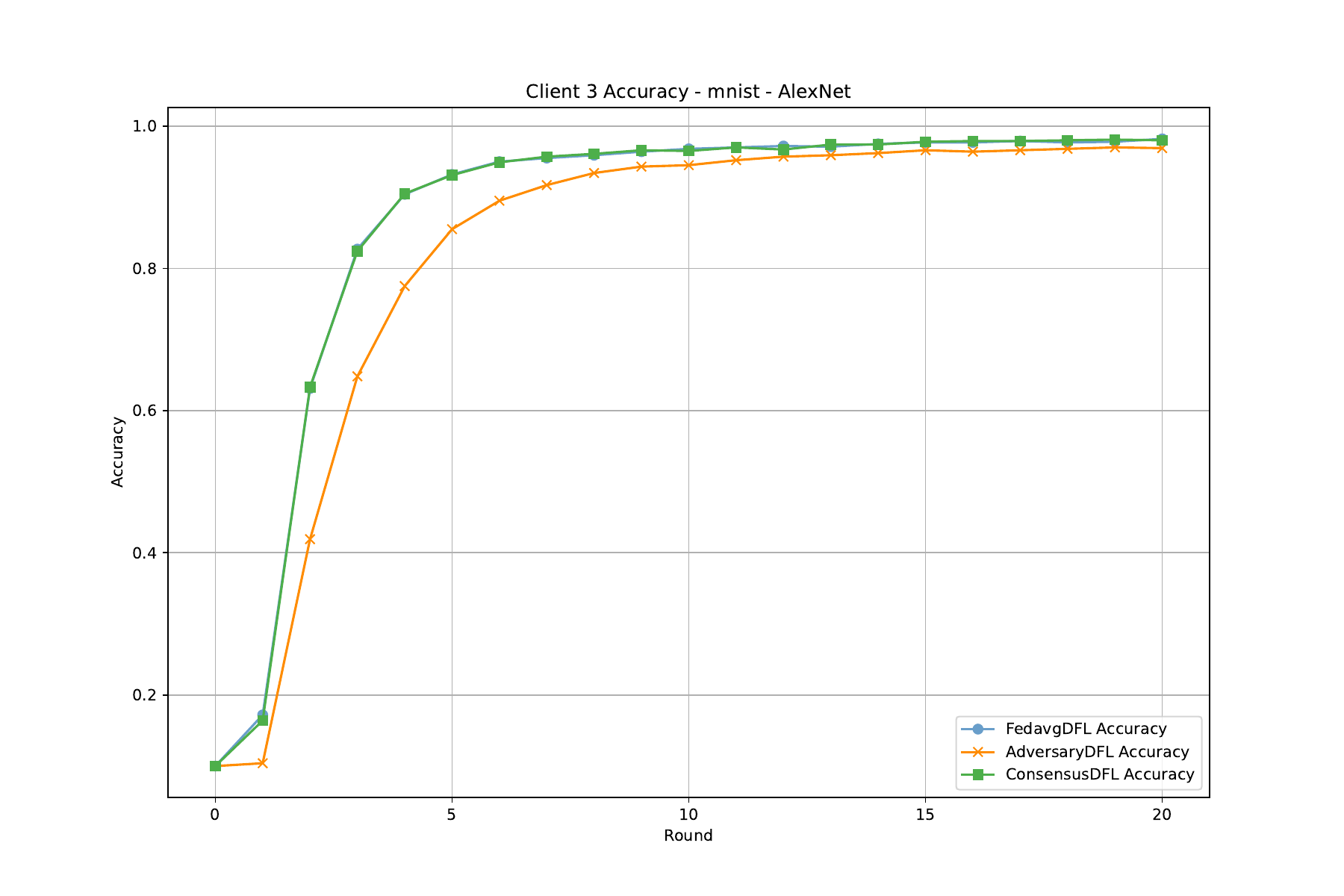}
        \end{tabular}
    }
    \vspace{0em} 
    \subfloat[Fashion-MNIST+AlexNet]{
        \begin{tabular}{cccc}
            \includegraphics[width=0.25\textwidth,trim=70 30 70 50,clip]{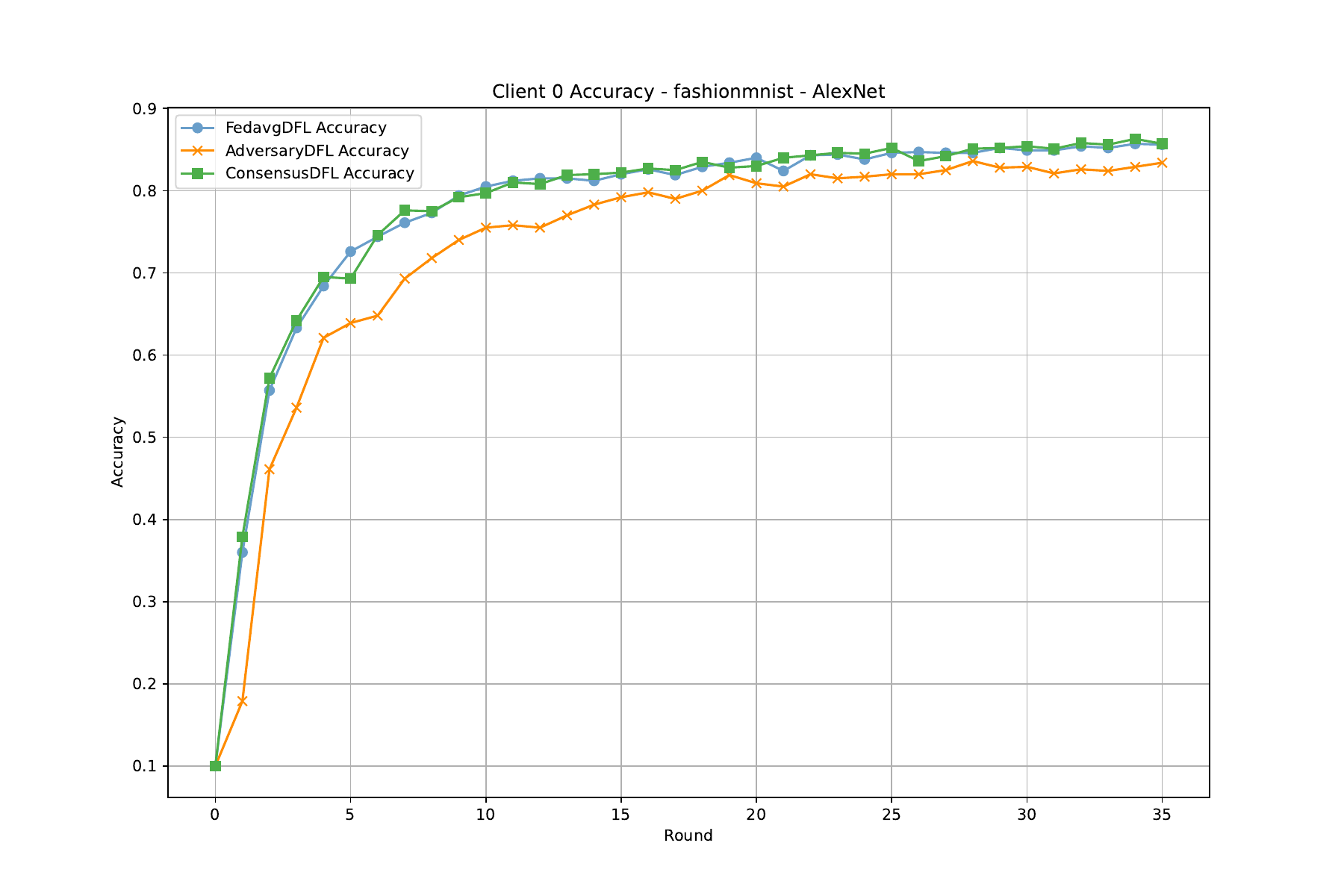} &
            \includegraphics[width=0.25\textwidth,trim=70 30 70 50,clip]{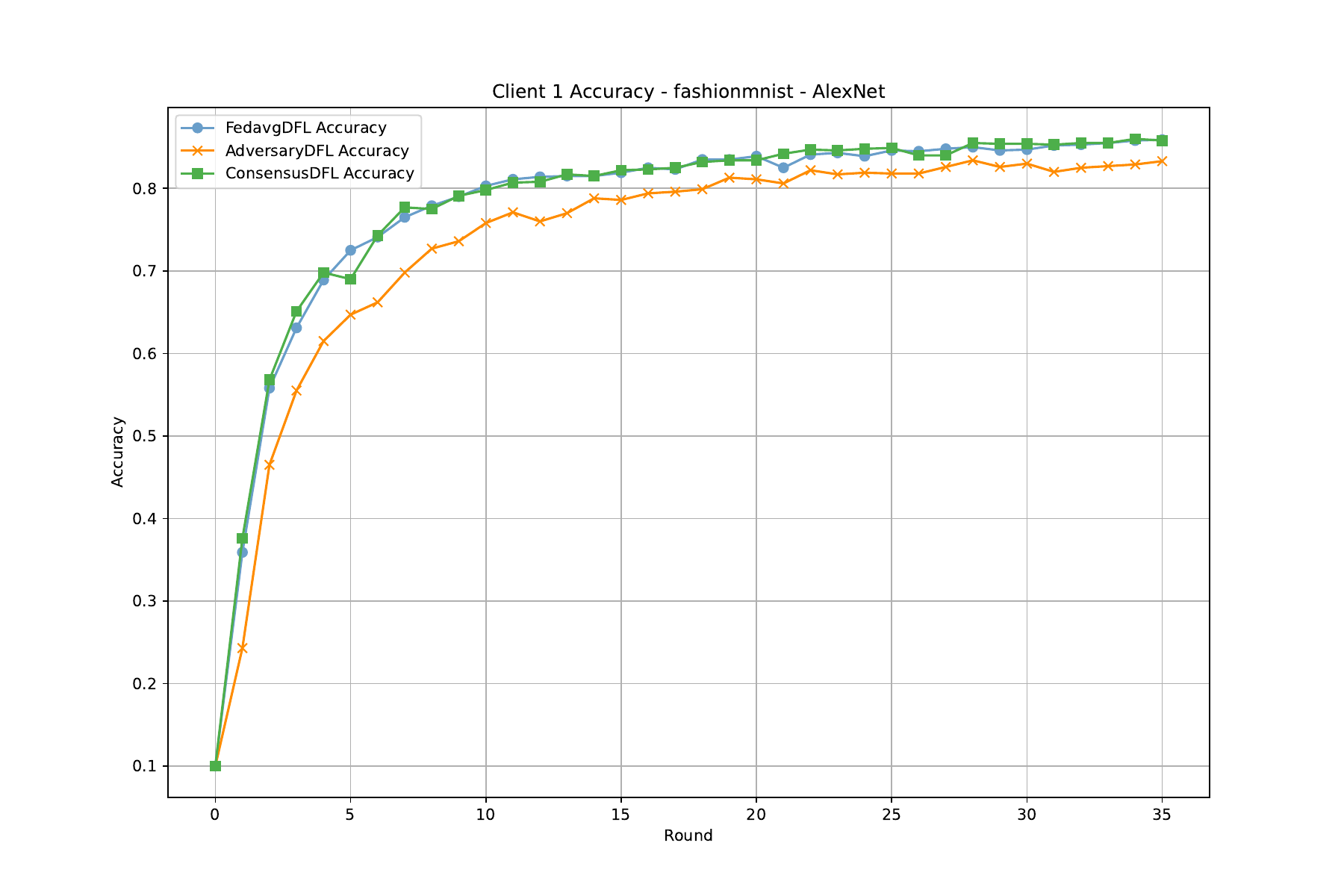} &
            \includegraphics[width=0.25\textwidth,trim=70 30 70 50,clip]{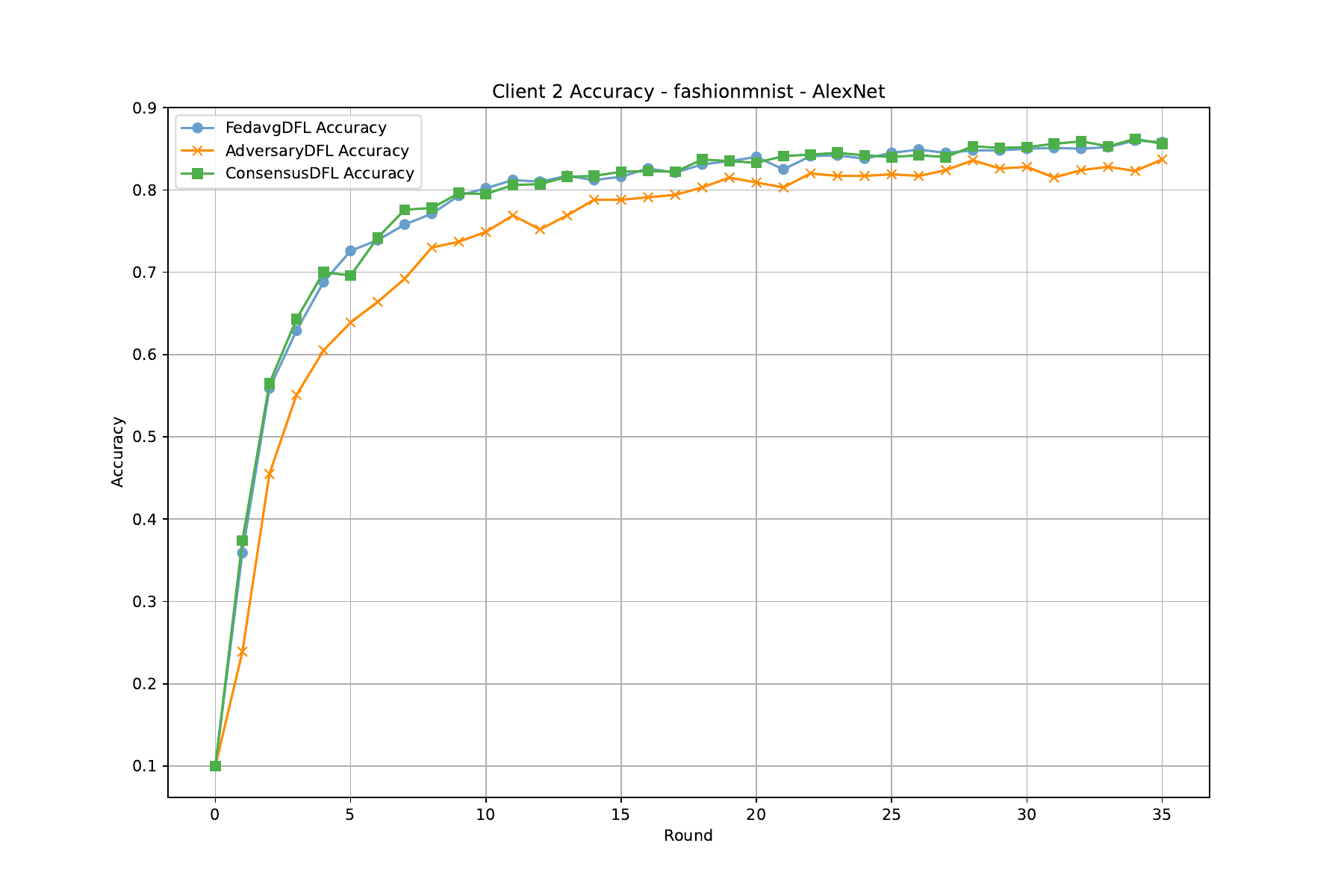} &
            \includegraphics[width=0.25\textwidth,trim=70 30 70 50,clip]{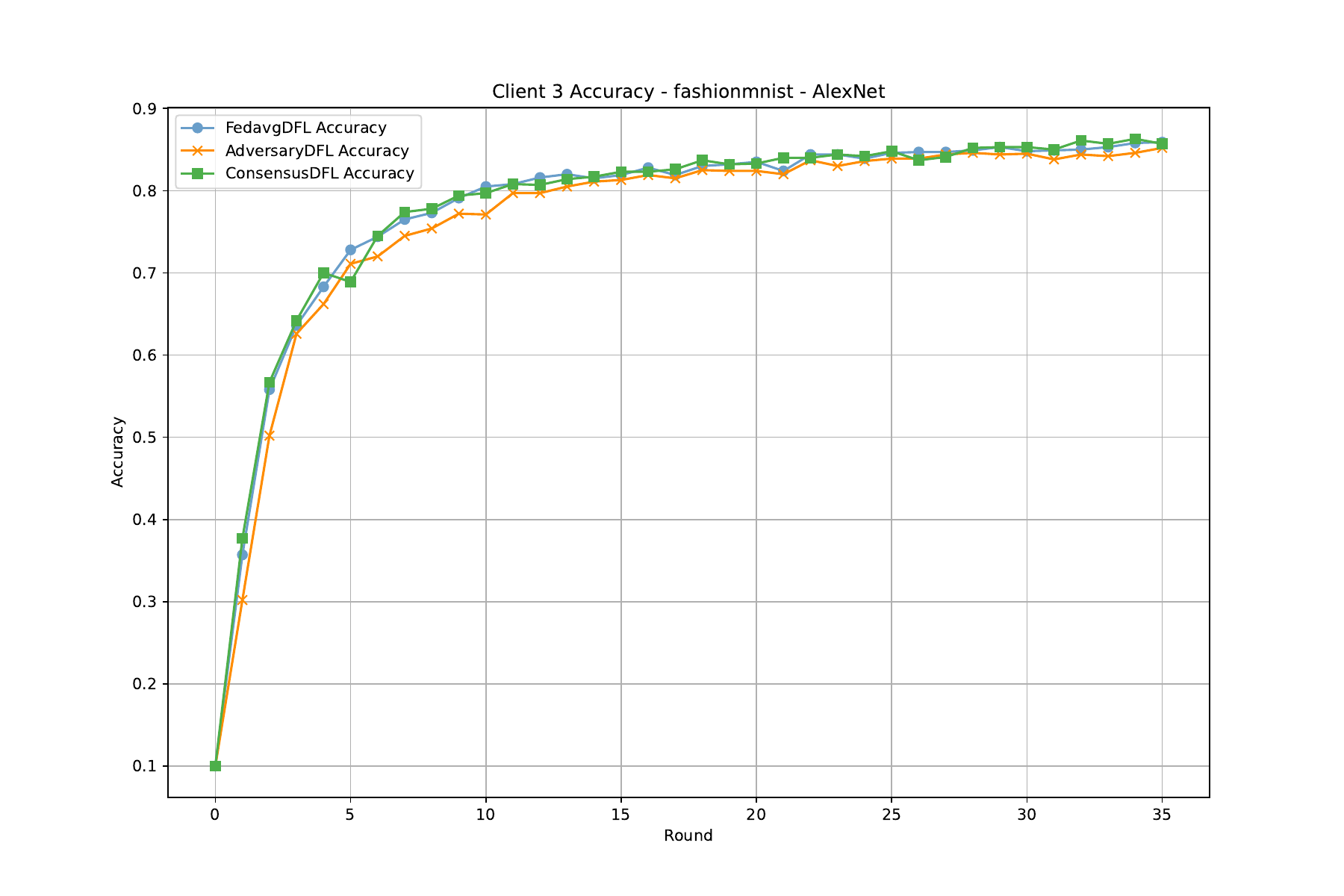}
        \end{tabular}
    }
    \vspace{0em} 
    \subfloat[CIFAR10+AlexNet]{
        \begin{tabular}{cccc}
            \includegraphics[width=0.25\textwidth,trim=70 30 70 50,clip]{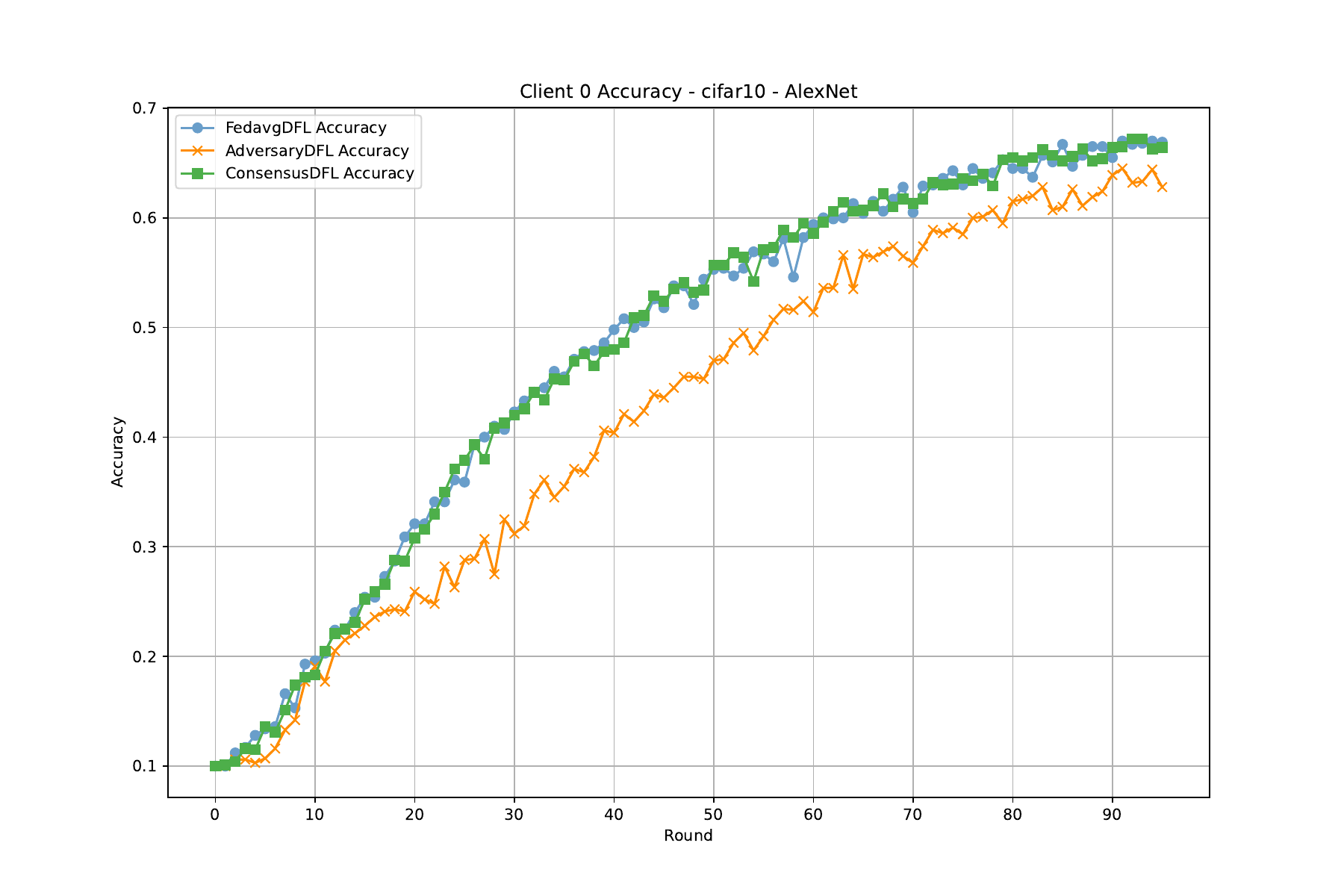} &
            \includegraphics[width=0.25\textwidth,trim=70 30 70 50,clip]{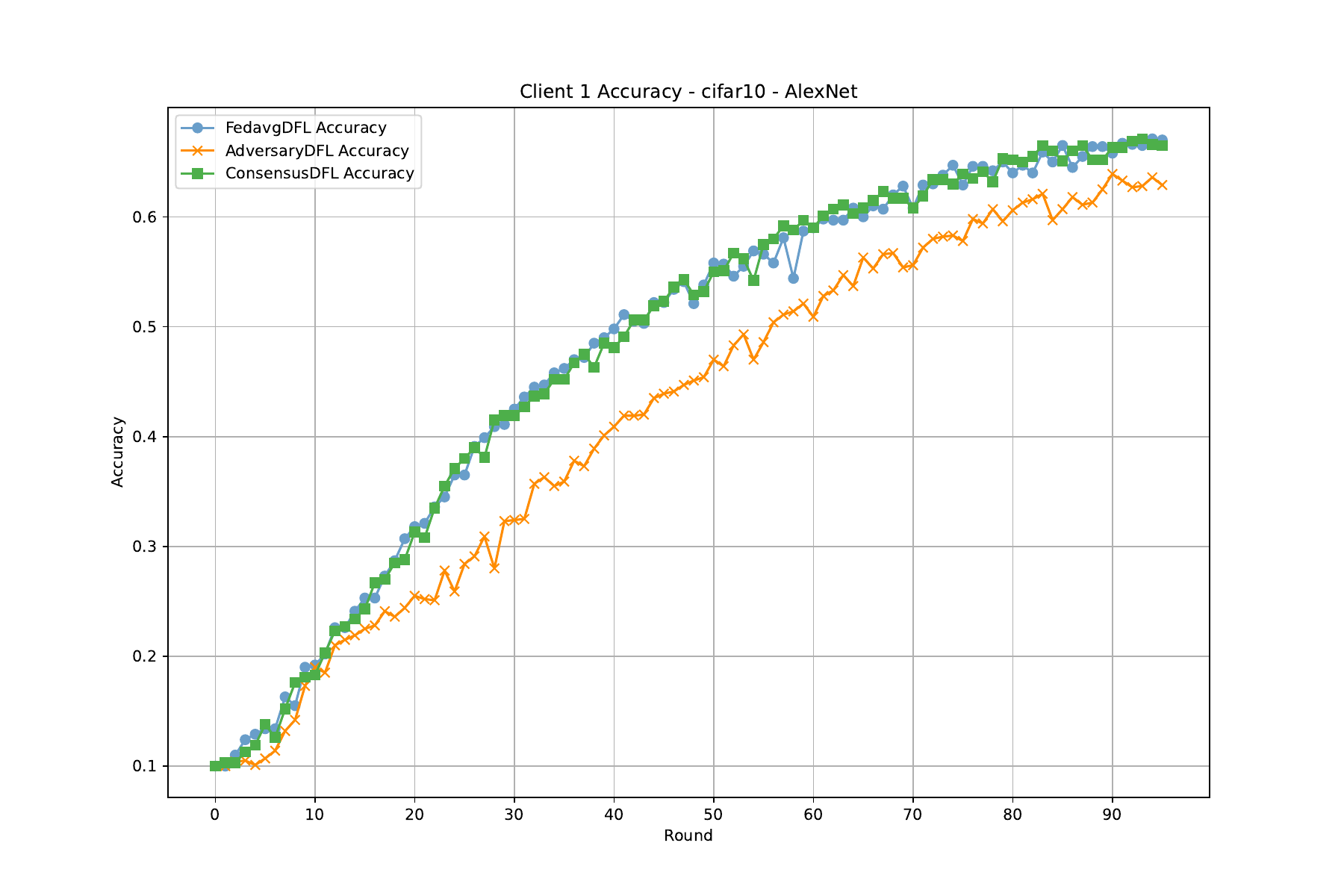} &
            \includegraphics[width=0.25\textwidth,trim=70 30 70 50,clip]{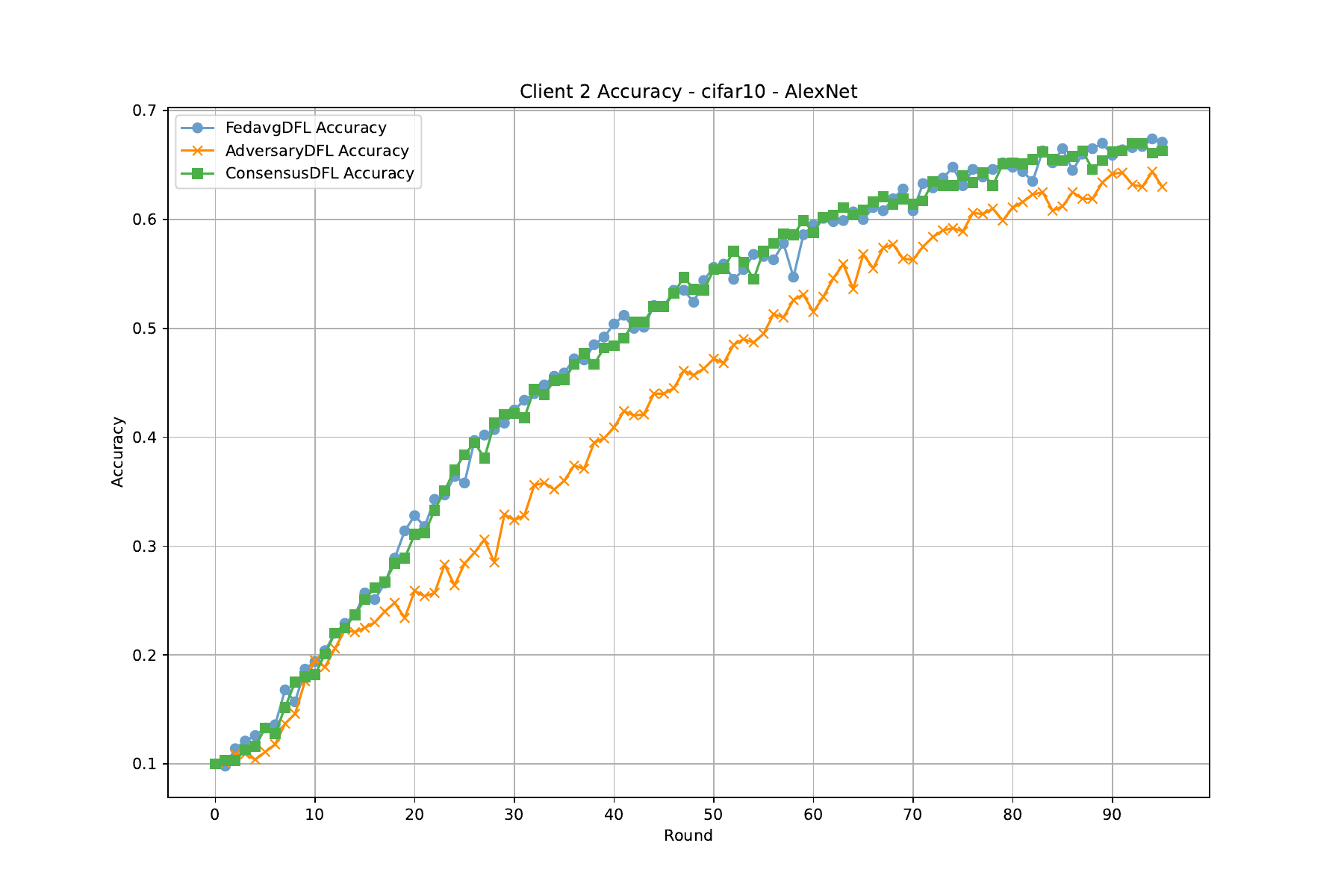} &
            \includegraphics[width=0.25\textwidth,trim=70 30 70 50,clip]{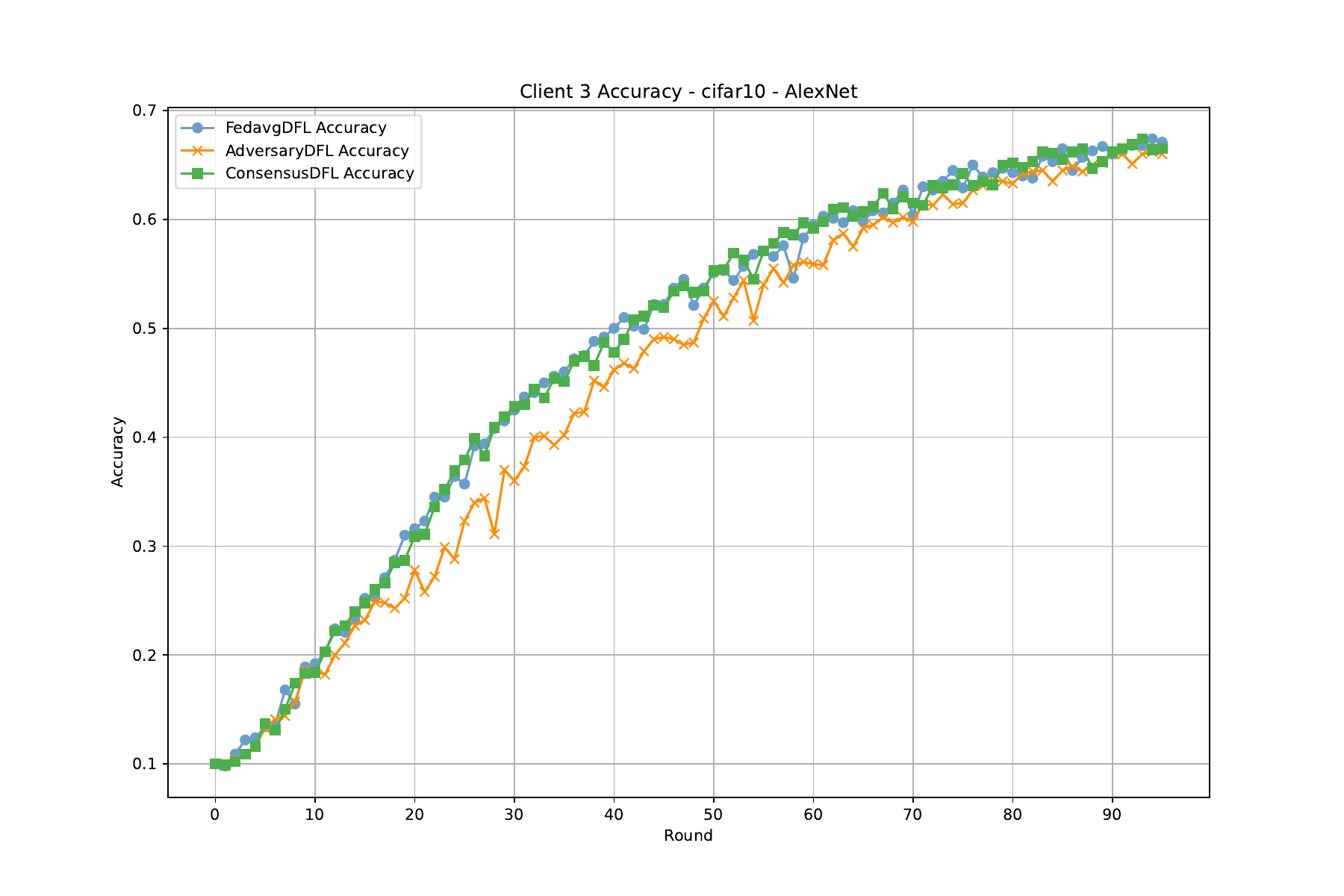}
        \end{tabular}
    }

    \caption{Accuracy Results using AlexNet model.}
    \label{fig:alexnet_results}
\end{figure*}

\begin{table*}[!t]
  \centering
  \caption{Experimental Results on Different Models and Datasets}
  \label{tab:exp_results}
  \renewcommand{\arraystretch}{1}
  \setlength{\tabcolsep}{8pt}
  \begin{tabular}{>{\centering\arraybackslash}m{2cm} >{\centering\arraybackslash}m{2.5cm} 
                  >{\centering\arraybackslash}m{1.2cm} >{\centering\arraybackslash}m{1.6cm} 
                  >{\centering\arraybackslash}m{1.2cm} >{\centering\arraybackslash}m{1.6cm} 
                  >{\centering\arraybackslash}m{1.6cm} >{\centering\arraybackslash}m{1.6cm}}
    \toprule
    \textbf{Model} & \textbf{Dataset} 
    & \multicolumn{2}{c}{\textbf{FedAvg}} 
    & \multicolumn{2}{c}{\textbf{ACuMPA}} 
    & \multicolumn{2}{c}{\textbf{ACuMPA + ByFTVeS}} \\
    \cmidrule(lr){3-4} \cmidrule(lr){5-6} \cmidrule(lr){7-8}
    & & Acc (\%) & IT (round) & Acc (\%) & IT (round) & Acc (\%) & IT (round) \\
    \midrule
    \multirow{3}{*}{CNN}
        & MNIST         & 98.6 & 30 & 98.6 & 36 & 98.9 & 30 \\
        & Fashion-MNIST & 86.9 & 13 & 84.6 & 22 & 87.0 & 12 \\
        & CIFAR10       & 63.7 & 66 & 60.8 & 115 & 63.8 & 69 \\
    \midrule
    \multirow{3}{*}{ResNet}
        & MNIST         & 99.2 & 8 & 99.0 & 10 & 99.1 & 9 \\
        & Fashion-MNIST & 90.5 & 23 & 89.4 & 48 & 90.6 & 22 \\
        & CIFAR10       & 70.5 & 24 & 67.7 & 43 & 70.5 & 27 \\
    \midrule
    \multirow{3}{*}{AlexNet}
        & MNIST         & 98.9 & 19 & 98.8 & 21 & 99.0 & 18 \\
        & Fashion-MNIST & 88.3 & 28 & 86.5 & 48 & 88.3 & 25 \\
        & CIFAR10       & 68.5 & 79 & 66.5 & 102 & 68.8 & 79 \\
    \bottomrule
  \end{tabular}
  \footnotesize{Note: The IT accuracy for each model on different datasets is as follows: All models on MNIST achieve 98\%, CNN on Fashion-MNIST reaches 80\%, CNN on CIFAR10 achieves 60\%, ResNet on Fashion-MNIST reaches 89\%, ResNet on CIFAR10 achieves 67\%, AlexNet on Fashion-MNIST reaches 85\%, and AlexNet on CIFAR10 achieves 65\%.}
\end{table*}

To evaluate the effectiveness of our proposed schemes in a distributed machine learning setting, we conduct nine experiments. All experimental results are summarized in Table~\ref{tab:exp_results}.

\textbf{Effectiveness of ACuMPA.}
To assess the impact of the proposed ACuMPA attack scheme, we compare it with the standard FedAvg method. As shown in Table~\ref{tab:exp_results}, the classification accuracy of benign participants under ACuMPA is consistently lower than that under FedAvg, indicating that ACuMPA successfully degrades the global model's performance.

Regarding convergence efficiency, we analyze the inference time—defined as the number of training rounds required to reach a given accuracy threshold. As illustrated in Figure~\ref{fig:cnn_results}, Figure~\ref{fig:resnet_results}, and Figure~\ref{fig:alexnet_results}, models trained with ACuMPA require more rounds to achieve the same accuracy compared to those trained with FedAvg. This suggests that ACuMPA not only impairs model performance but also hinders training efficiency.

Additionally, we observe that the malicious participant is less affected by the attack than the benign ones, which aligns with the adversary’s motivation to sabotage global training while preserving their own model utility.

The proposed ACuMPA scheme can effectively compromise the performance of all participants in a distributed learning system. Compared to the baseline FedAvg setting without attack, FedAvg integrated with ACuMPA leads to lower model accuracy and increased inference time. Furthermore, the degradation experienced by benign participants is greater than that observed for the malicious participant, which is consistent with the intended objectives of the adversary.

\textbf{Effectiveness of ByFTVeS.}
As demonstrated above, the ACuMPA scheme effectively compromises the performance of participants in a distributed machine learning system. To defend against this threat, we introduce ByFTVeS as a mitigation mechanism and evaluate its effectiveness when applied alongside ACuMPA.

As shown in Table~\ref{tab:exp_results}, the model accuracy of benign participants under the ACuMPA + ByFTVeS setting is approximately the same as that under the standard FedAvg setting without attack. This indicates that ByFTVeS effectively neutralizes the malicious impact of ACuMPA while maintaining model utility.

In terms of inference time, as illustrated in Figure~\ref{fig:cnn_results}, Figure~\ref{fig:resnet_results}, and Figure~\ref{fig:alexnet_results}, the number of training rounds required to reach a predefined accuracy threshold under the ACuMPA + ByFTVeS setting is nearly identical to that under FedAvg. This further validates the ability of ByFTVeS to preserve training efficiency while defending against adversarial behavior.


\textbf{Summary.}
The experimental results in the distributed machine learning setting demonstrate that our proposed ACuMPA scheme significantly degrades both model accuracy and convergence efficiency compared to the baseline FedAvg. Furthermore, the malicious participant remains largely unaffected, which aligns with its adversarial objective. To counteract this threat, we propose the ByFTVeS defense mechanism. Experimental results show that ByFTVeS effectively mitigates the impact of ACuMPA, preserving both model accuracy and training efficiency, achieving performance comparable to the standard FedAvg. These findings are consistent with our theoretical analysis and confirm the practical effectiveness of both the attack and defense mechanisms.

\subsection{Evaluation on MPC}
\subsubsection{Experimental Settings}
To evaluate the scalability of our proposed ByFTVeS scheme, we extend ByFTVeS to Secure Multi-Party Computation(MPC) scenario.

{\bf Setup.} We consider a distributed MPC framework involving three participants, each participant(denoted as $p_i$) have a local sensitive dataset(denoted as $s_i$) and they collaborate to privacy-preserving calculate a function $f(s_0,s_1,s_2)$. This framwork is equipped with our proposed ByFTVeS scheme to keep sensitivity.

{\bf Setting.} We implement a prototype of the proposed schemes using PyTorch and Python 3.8. All experiments are conducted on a server running Ubuntu 22.04, equipped with a 16-core Intel(R) Xeon(R) Gold 6430 CPU, 120 GB of RAM, and an NVIDIA GeForce RTX 4090 GPU.

{\bf Datasets.} To evaluate our proposed schemes in a real-world application, we use the Carbon emissions data from five districts in a city from China. This dataset contains simulated private data from power supply(denoted as $E_{pow}$), energy(denoted as $E_{ene}$) and ecology(denoted as $E_{eco}$), and includes 1,652 data.

{\bf Metrics.} We evaluate the performance of the MPC protocol using a composite energy function defined as $E = E_{pow} + E_{ene} + E_{eco}$, which supports privacy-preserving summation. To assess the efficiency of the MPC protocol, we use computation time as the primary evaluation metric.

{\bf Baseline.} To evaluate the computational efficiency of our scheme, we compare it against MASCOT, a widely adopted state-of-the-art (SOTA) MPC protocol known for its efficiency in secure two-party and multi-party computation. MASCOT employs preprocessed oblivious transfer (OT) and authenticated shares to enable efficient arithmetic operations over finite fields, making it a strong benchmark for performance comparison.

\subsubsection{Experimental Results}
We conduct experiments to evaluate the effectiveness of our proposed ByFTVeS scheme in the MPC application. All experimental results are summarized in Table~\ref{tab:comp_time}.

\begin{table}[!t]
  \centering
  \caption{Computation Time Comparison between ByFTVeS and MASCOT (in seconds)}
  \label{tab:comp_time}
  \renewcommand{\arraystretch}{1}
  \setlength{\tabcolsep}{5pt}
  \begin{tabular}{l|c|c|l|c|c}
    \toprule
   \textbf{Dataset}  & \textbf{EyFTVeS} & \textbf{MASCOT} &  \textbf{Dataset}  & \textbf{EyFTVeS} & \textbf{MASCOT}\\
    \midrule
     District A & 0.0114 & 655& 
    District B & 0.0086 & 1553 \\
    District C & 0.0040 & 520 & 
    District D & 0.0221 & 135 \\
    District E & 0.0045 & 194 &
    District F & 0.0173 & 1208 \\
    \bottomrule
  \end{tabular}
\end{table}

\section{Conclusion}
In this paper, we propose the ASDP strategy, enabling adversaries to generate inconsistent shares for victims. Consequently, a victim's  model becomes compromised when attacked by ACuMPA—a novel poisoning attack built upon ASDP. To validate the effectiveness of both ASDP and ACuMPA, we conduct extensive distributed privacy-preserving machine learning experiments and provide theoretical proofs. 

To counteract the threat, we propose EByFTVeS scheme that force honest participant to commit their shares before using them to calculate the average. Experiments show that the EByFTVeS scheme effectively defends against ACuMPA attacks. We further provide formal analyses of EByFTVeS's validity, consistency, liveness. Furthermore, we evaluated our scheme  on a real-world MPC application, and outperforms MASCOT (the state-of-the-art MPC protocol) in efficiency.


\bibliographystyle{IEEEtran}
\bibliography{IEEEabrv,ref.bib}

\newpage
\vspace{11pt}
\vfill
\end{document}